\documentclass[fleqn,usenatbib]{mnras}
\usepackage{fix-cm}
\usepackage{newtxtext,newtxmath}

\usepackage[T1]{fontenc}

\usepackage{array}

\DeclareRobustCommand{\VAN}[3]{#2}
\let\VANthebibliography\thebibliography
\def\thebibliography{\DeclareRobustCommand{\VAN}[3]{##3}\VANthebibliography}

\usepackage{ulem}

\newcommand{\feh}{[\rm Fe/H]}
\newcommand{\afe}{[\rm \alpha/Fe]}
\usepackage{graphicx}	
\usepackage{amsmath}	
\usepackage{hyperref}

\title[HST-JWST analysis of Draco\,II]{A combined \textit{HST} and \textit{JWST} analysis of the stellar population in Draco\,II}

\author[Axelle C. Raffin et al.]{Axelle C. Raffin$^{1}$\thanks{E-mail: axelle.raffin@anu.edu.au},
Giacomo Cordoni$^{1}$,
Helmut Jerjen$^{1}$,
Gary S. Da Costa$^{1}$,
Maria V. Legnardi$^{2}$,
\newauthor
Anna F. Marino$^{3,4}$,
Antonino P. Milone$^{2,3}$,
and Fabrizio Muratore$^{2}$\\
\\
$^{1}$Research School of Astronomy and Astrophysics, The Australian National University, Mt Stromlo Observatory, Canberra, ACT 2611, Australia\\
$^{2}$Dipartimento di Fisica e Astronomia ‘Galileo Galilei’, Università di Padova, Vicolo dell’Osservatorio 3, Padova, IT-35122, Italy\\
$^{3}$INAF -- Osservatorio Astronomico di Padova, Vicolo dell’Osservatorio 5, Padova, IT-35122, Italy\\
$^{4}$INAF -- Osservatorio Astrofisico di Arcetri, Largo Enrico Fermi 5, Firenze, IT-50125, Italy
}

\date{Accepted 2026 October 19. Received 2026 October 16; in original form 2026 June 2}

\pubyear{\the\year{}}

\begin{document}
\label{firstpage}
\pagerange{\pageref{firstpage}--\pageref{lastpage}}
\maketitle

\begin{abstract}
Draco\,II (Dra\,II) is one of the faintest and most metal-poor Milky Way satellites known, yet its nature as either an ultra-faint dwarf galaxy or an extended globular cluster remains debated. We perform a photometric analysis of the stellar population  of Dra\,II reducing deep multi-band observations from the \textit{Hubble Space Telescope} (\textit{HST}) and the \textit{James Webb Space Telescope} (\textit{JWST}). We select Dra\,II member stars based on their proper motions and photometric quality criteria. For the first time, optical and infrared photometry are combined to produce a stellar catalogue reaching 50\,percent completeness at $F814W = 26.5$\,mag (corresponding to the absolute magnitude $M_{F814W} = 10.1$\,mag). 
The colour-magnitude diagram confirms that Dra\,II hosts an old stellar population with an age of $13.5_{-1.5}^{+1.7}$\,Gyr at a distance of $19.2\pm0.9$\,kpc. We measure an $\alpha$-enhanced stellar population with a low mean metallicity of $\feh_{CaHK} = -2.38 \pm 0.17$\,dex and confirm the presence of an intrinsic metallicity spread of $\sigma_{\feh} = 0.85 \pm 0.13$\,dex. 
We identify 10 unresolved main-sequence binary stars in the HST catalogue with mass ratios $q > 0.5$ in the mass range $0.43<M_1/M_\odot<0.71$, corresponding to a binary fraction of $f_{bin}^{q>0.5} = 0.21 \pm 0.07$. This value is higher than those typically observed in globular clusters and lies near the upper range measured in open clusters. The stellar mass function is best modelled by a single power law with a slope of $1.53\pm0.26$ over the mass range $0.13$--$0.77\ M_\odot$. We find no evidence of mass segregation. Consistent results are obtained independently from the JWST catalogue. Combining these findings strongly suggests that Draco\,II is best classified as an ultra-faint dwarf galaxy rather than a globular cluster.
\end{abstract}




\section{Introduction}

Large sky surveys like the Sloan Digital Sky Survey \citep[SDSS;][]{York:2000}, the Dark Energy Survey \citep[DES;][]{DES:2005} or the Panoramic Survey Telescope and Rapid Response System 1 3$\pi$ survey\footnote{https://outerspace.stsci.edu/display/PANSTARRS/} \citep[Pan-STARRS 1 3$\pi$;][]{Chambers:2016} uncovered an abundance of elusive Milky Way satellites \citep{Willman:2005, Belokurov:2007, Kim:2015}. Among these systems, ultra-faint dwarf galaxies \citep[UFDs;][]{Simon:2019} are particularly important to study galaxy formation and dark matter. UFDs are extreme environments: they are the \textit{oldest, most dark-matter dominated, most metal-poor, and least chemically evolved stellar systems known} \citep{Simon:2019}, making them valuable laboratories for understanding the early stages of galaxy formation and chemical enrichment in the Universe \citep{Kroupa:2010, Bullock:2017}. Indeed, as dwarf galaxies (DG) are embedded in dark matter halos \citep{Kauffmann:1993, Strigari:2008, Tolstoy:2009, Galaxy} they provide key constraints on the nature of dark matter and the low-mass end of galaxy formation within $\Lambda$CDM cosmology \citep{Strigari:2018, Simon:2019}.

Characterising the stellar population of UFDs remains difficult due to their extremely low luminosities and sparse stellar populations. Deep photometric observations with the Hubble Space Telescope (HST) and the James Webb Space Telescope (JWST) have therefore become essential to resolve their stellar populations and constrain their star formation histories, metallicity distributions, and stellar mass functions \citep{Geha:2013, Weisz:2017, Simon:2019, Fu:2023, Milone:2023}. Combining optical and infrared observations is particularly powerful, as it enables improved sensitivity to low-mass stars with the infrared regime and to metallicity variations with blue and ultraviolet filters, while simultaneously constraining stellar population properties across a broad wavelength range.

The system Draco\,II (Dra\,II) is a particularly challenging Milky Way satellite. Discovered by \citet{Laevens:2015} in the Pan-STARRS 1 3$\pi$ survey and located at $\alpha_{2000} = 15^\mathrm{h}52^\mathrm{m}47.60^\mathrm{s}$, $\delta_{2000} = +64^\circ33'55.0''$ \citep{Weisz:2023}, Dra\,II is both extremely faint \citep[$M_V = -2.9\pm0.8$;][]{Laevens:2015} and compact \citep[$r_h = 19_{-2.6}^{+4.5}$\, pc;][]{Longeard:2018}. Despite multiple observations, its true nature remains uncertain: Dra\,II may either be an ultra-faint dwarf spheroidal galaxy or an extended globular cluster (GC). Distinguishing between UFDs and star clusters is inherently difficult due their similar characteristics (size, luminosity, morphology), particularly at extremely faint luminosities where the properties of both systems overlap. Determining the nature of Dra\,II is nevertheless important, as the presence of a dark matter halo would make Dra\,II an excellent laboratory for the study of dark matter and $\Lambda$CDM cosmology \citep{Kroupa:2010, Bullock:2017}.

Previous studies have shown that Dra\,II is one of the oldest, most metal-poor satellites of the Milky Way, with a mean metallicity of $\feh = -2.72_{-0.11}^{+0.12}$\,dex \citep{Fu:2023}. All previous works concluded that Dra\,II has an $\alpha$-enhanced stellar population \citep{Longeard:2018, Fu:2023, Weisz:2023}. The low metallicity combined with the $\alpha$-enhancement suggests an old stellar population; indeed, isochrone fitting yields an age of $13.5 \pm 0.5$ Gyr \citep{Longeard:2018}. However, several questions remain unresolved. In particular, the existence of a metallicity spread, a hallmark of chemical self-enrichment in DGs, remains debated \citep{Martin:2016, Longeard:2018, Fu:2023, Cerny:2026}. Similarly, the presence or absence of mass segregation has also been investigated in Dra\,II, without a clear consensus reached \citep{Longeard:2018, Baumgardt:2022}. These uncertainties are partly driven by the limited number of observable stars and the faintness of the system. A compilation of previous results is presented in Table~\ref{tab:draII_params}.

\begin{table*}
\centering
\label{tab:draII_params}
\caption{Physical parameters of Dra\,II from previous studies. The values in bold indicate results from spectroscopic data (the results of \citet{Martin:2016, Cerny:2026}, and some of the results of \citet{Longeard:2018}) while the other results come from photometric data.}


\resizebox{\textwidth}{!}{
\begin{tabular}{lcccccc}
\hline\hline
Parameter & Laevens et al. (2015) & Martin et al. (2016) & Longeard et al. (2018) & Fu et al. (2023) & Weisz et al. (2023) & Cerny et al. (2026) \\ [1.2ex]
\hline
d [kpc] & $20\pm0.3$ & ... & $21.5 \pm 0.4$ & ... & ... & ... \\[1.2ex]
$M_V$ [mag] & $-2.9 \pm 0.8$ & ... & $-0.8_{-1.0}^{+0.4}$ & ... & ... & ... \\[1.2ex]
$r_h$ [arcmin] & $2.7_{-0.8}^{+1.0}$ & ... & $3.0_{-0.5}^{+0.7}$ & ... & ... & ... \\[1.2ex]
$r_h$ [pc] & $19_{-6}^{+8}$ & ... & $19_{-2.6}^{+4.5}$ & ... & ... & ... \\[1.2ex]
$\langle v_r \rangle$ [km s$^{-1}$] & ... & $\boldsymbol{-347.6_{-1.8}^{+1.7}}$ & $\boldsymbol{-342.5_{-1.2}^{+1.1}}$ & ... & ... & $\boldsymbol{-342.4 \pm 0.7}$ \\[1.2ex]
$\sigma_{v_r}$ [km s$^{-1}$] & ... & $\boldsymbol{2.9 \pm 2.1}$ & $\boldsymbol{< 5.9}$ & ... & ... & $\boldsymbol{< 2.6}$ \\[1.2ex]
$\log(M_{1/2}/M_\odot)$ & ... & $\boldsymbol{5.5_{-0.6}^{+0.4}}$ & ... & ... & ... & ... \\[1.2ex]
$\feh_{CMD}$ [dex] & $\sim -2.2$ & ... & $-2.4 \pm 0.05$ & ... & between $-2.0$ and $-1.4$ & ... \\[1.2ex]
$\feh_{CaII}$ [dex] & ... & $\boldsymbol{< -2.1}$ & ... & ... & ... & ... \\[1.2ex]
$\feh_{spectro}$ [dex] & ... & ... & $\boldsymbol{-2.43_{-0.82}^{+0.41}}$ & ... & ... & $\boldsymbol{-2.94_{-0.39}^{+0.32}}$ \\[1.2ex]
$\feh_{CaHK}$ [dex] & ... & ... & $-2.7 \pm 0.1$ & $-2.72_{-0.11}^{+0.10}$ & ... & ... \\[1.2ex]
$\sigma_{\feh}$ [dex] & ... & \textbf{suggested} & $< 0.24$ & $0.40_{-0.12}^{+0.12}$ & $\sim0.5$ & $\boldsymbol{0.57 \pm 0.27}$ \\[1.2ex]
$\afe$ [dex] & ... & ... & $0.6$ & $0.4$ & $\alpha$-enhanced & ... \\[1.2ex]
Age [Gyr] & $\sim12$ & ... & $13.5 \pm 0.5$ & ... & $\sim13$ & ... \\[1.2ex]
\hline
\end{tabular}
}
\end{table*}

In this work, we analyse deep multi-band HST and JWST photometry to characterise the stellar population of Dra\,II. By combining optical and infrared photometry, we investigate several of Dra\,II's properties, including age, metallicity, stellar mass function, mass segregation, and the presence of unresolved binary stars. Our goal is to assess whether these properties are more consistent with those of a globular cluster or an ultra-faint dwarf galaxy. The paper is organised as follows: in Section~\ref{sec:observations} we present the datasets and data reduction pipeline used in this work, Section~\ref{sec:results} presents the analyses and results, while Section~\ref{sec:discussion} discusses their implications and summarises our conclusion.

\section{Observations and Data Reduction}
\label{sec:observations}

\subsection{Observations}

\begin{figure}
    \includegraphics[width=\columnwidth]{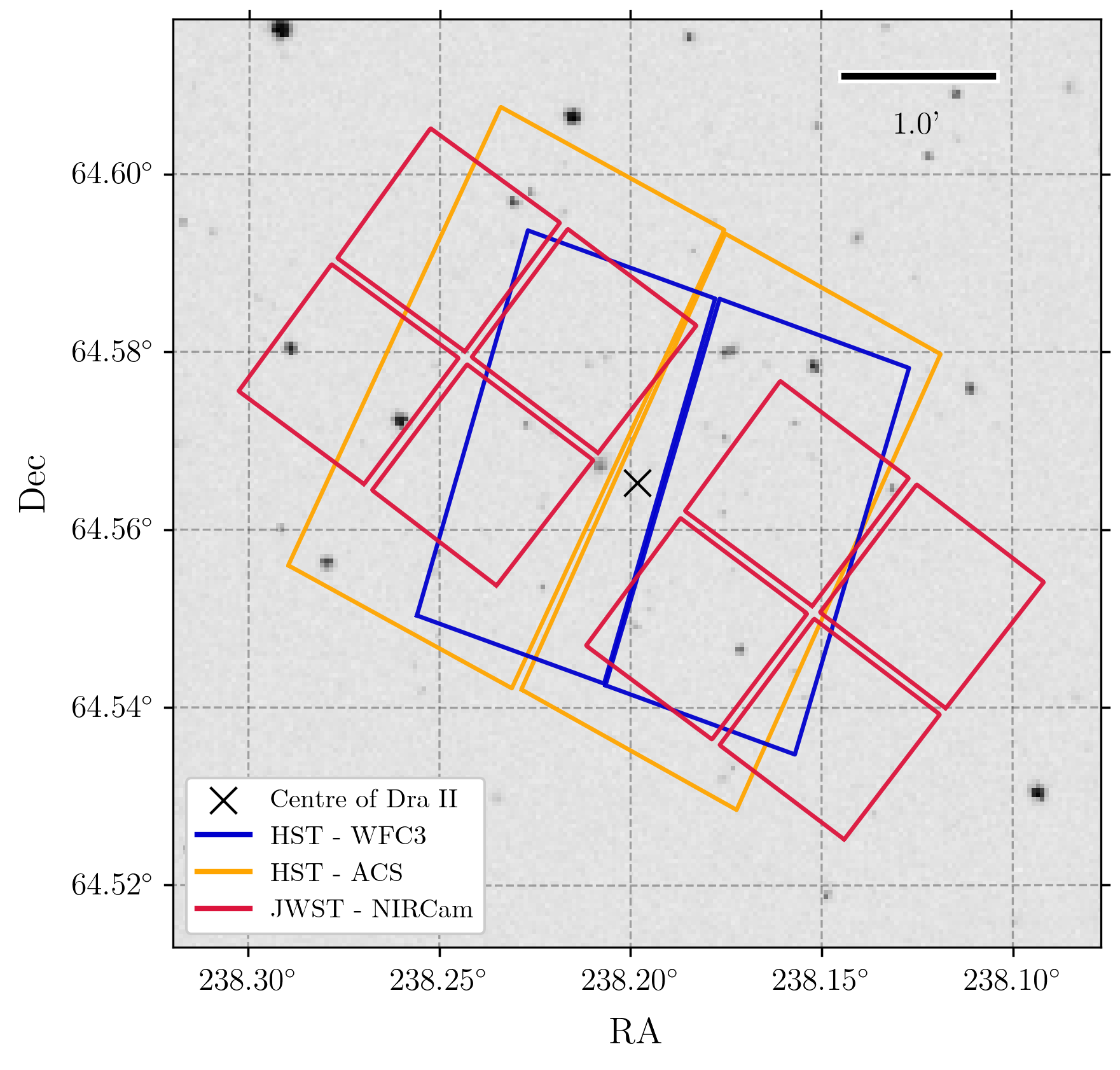}
    \caption{Footprint of the HST (blue and orange) and JWST (red) fields relative to Dra\,II's nominal centre ($\alpha_{2000} = 15^\mathrm{h}52^\mathrm{m}47.60^\mathrm{s}$, $\delta_{2000} = +64^\circ33'55.0''$). The scale bar corresponds to 1 arcmin (scale: 5.6 pc/arcmin); the half-light radius of Dra\,II is $r_h \approx 3$ arcmin.}
    \label{fig:HST_JWST_FoVs}
\end{figure}

We analyse imaging data from HST (proposal IDs: $14734$, $15901$ and $17473$; PIs: Kallivayalil, Weisz, and  Warfield, respectively) and JWST (proposal IDs: $1334$ and $4570$; PIs: Weisz, and Warfield) for Draco\,II, publicly available on the Mikulski Archive for Space Telescopes\footnote{https://mast.stsci.edu/search/ui/} (MAST). HST images were obtained with two instruments: the Advanced Camera for Surveys (ACS) Wide Field Camera channel (WFC) and the Wide Field Camera 3 (WFC3) ultraviolet-visible channel (UVIS). JWST images were acquired with the Near-InfraRed Camera (NIRCam). 
Table~\ref{tab:imagelist} lists all the HST and JWST datasets we reduced. For several HST datasets, both short and long exposures are available. Due to the physical gap between HST’s two detectors, the observations were carefully dithered so that stars falling in the inter-chip gap in long exposures do so no more than once, ensuring spatial uniformity \citep{Anderson:2008}. The footprints of the observations used in this work are presented in Fig.~\ref{fig:HST_JWST_FoVs}.

\begin{table}
\centering
    \caption{
    HST and JWST datasets used for this study of Dra\,II, publicly available on MAST.}
        \begin{tabular}{cccc}
          \hline\hline           
            Dataset ID & Filter & Epoch & $t_{exp}$ (s)\\
                       (1) & (2) & (3) & (4)\\
          \hline 
                    \multicolumn{4}{c}{HST}\\\hline
            JDAN19010 & F814W & 02 Mar 2017 & 4662\\
            
            JDAN18010 & F606W & 06 Mar 2017 & 4662\\
            
            IE4L18030 & F475W & 23 Apr 2020 & 919\\
           
            IE4L18040 & F395N & 23 Apr 2020 & 8040\\
            
            JF9Q01010 & F814W & 06 Mar 2024 & 5362\\
            
            JF9Q03010 & F814W & 06 Mar 2024 & 5362\\
          \hline 
                  \multicolumn{4}{c}{JWST}\\\hline
            jw01334003001\_02101\_00001 & F090W & 03 Jul 2022 & 2952.612\\
           
            jw01334003001\_02101\_00002 & F090W & 03 Jul 2022 & 2952.612\\
            
            jw01334003001\_02101\_00003 & F090W & 03 Jul 2022 & 2952.612\\
            
            jw01334003001\_02101\_00004 & F090W & 03 Jul 2022 & 2952.612\\
            
            jw01334003001\_04101\_00001 & F150W & 04 Jul 2022 & 1470.937\\
            
            jw01334003001\_04101\_00002 & F150W & 04 Jul 2022 & 1470.937\\
            
            jw01334003001\_04101\_00003 & F150W & 04 Jul 2022 & 1470.937\\
            
            jw01334003001\_04101\_00004 & F150W & 04 Jul 2022 & 1470.937\\
          \hline 
          
         \end{tabular}

         \label{tab:imagelist}
\end{table}

\subsection{Data reduction}
We measure stellar fluxes and positions in the HST images using the \texttt{FORTRAN} program KS2 \citep{Anderson:2008}. The exquisite precision of KS2 relies on determining an accurate effective Point Spread Function \citep[ePSF;][]{Anderson:2000}, an empirical model describing the fraction of stellar light falling on each pixel. By iteratively refining the ePSF using dithered exposures of real stars, the model achieves high systematic accuracy for astrometry and photometry \citep[e.g.][]{Bellini:2017, Libralato:2022, 47Tuc}. KS2 then uses three different methods to measure stars. Method I is optimal for bright stars and derives magnitudes and positions of stars in each individual exposure using the best ePSF model, then averages them to improve the measurements. On the other hand, method II, which stacks all images before measuring fluxes and positions, performs better for faint sources by carrying out aperture photometry for each star on a 5 $\times$ 5 pixel raster. Finally, method III is optimal for crowded regions: similar to method II, it calculates the photometry within a circular aperture of radius 0.75 pixels. The astrometry obtained is further refined by combining the data with observations of Dra\,II in the Gaia DR3 \citep{GaiaCollab:2023}, which provides a more precise reference frame for the coordinates of the stars.

We apply a similar process to the JWST NIRCam observations, by deriving a grid of effective PSFs for stellar flux and magnitude measurements. We calibrate the photometry to the VEGA magnitude system using aperture photometry on the drizzled files and the photometric zero points for ACS\footnote{https://acszeropoints.stsci.edu/}, WFC3\footnote{https://www.stsci.edu/hst/instrumentation/wfc3/data-analysis/photometric-calibration} and NIRCam\footnote{https://jwst-docs.stsci.edu/jwst-near-infrared-camera/nircam-performance/nircam-absolute-flux-calibration-and-zeropoints}. 

Finally, we perform artificial star tests to assess the performance of the data reduction pipeline and to obtain the photometric uncertainties and detection completeness for the entire dataset, both expressed as functions of magnitude.

\subsection{Member selection}
We apply selection cuts to both source catalogues to separate Dra\,II member stars from foreground and background contaminants. For the HST data, three observing epochs spanning seven years (2017, 2020, and 2024) enable the measurement of the relative proper motions of stars. Images from each epoch are reduced independently, producing a separate astrometric and photometric catalogue per epoch. The first-epoch (2017) catalogue serves as the master reference frame, and we align subsequent epochs to it using six-parameter linear transformations \citep{Anderson:2006} determined from the F814W magnitude of stars. From these cross-matched catalogues, we derive the relative displacement of stars over the seven-years baseline and convert them to yearly relative proper motions. 

\begin{figure}
	\includegraphics[width=\columnwidth]{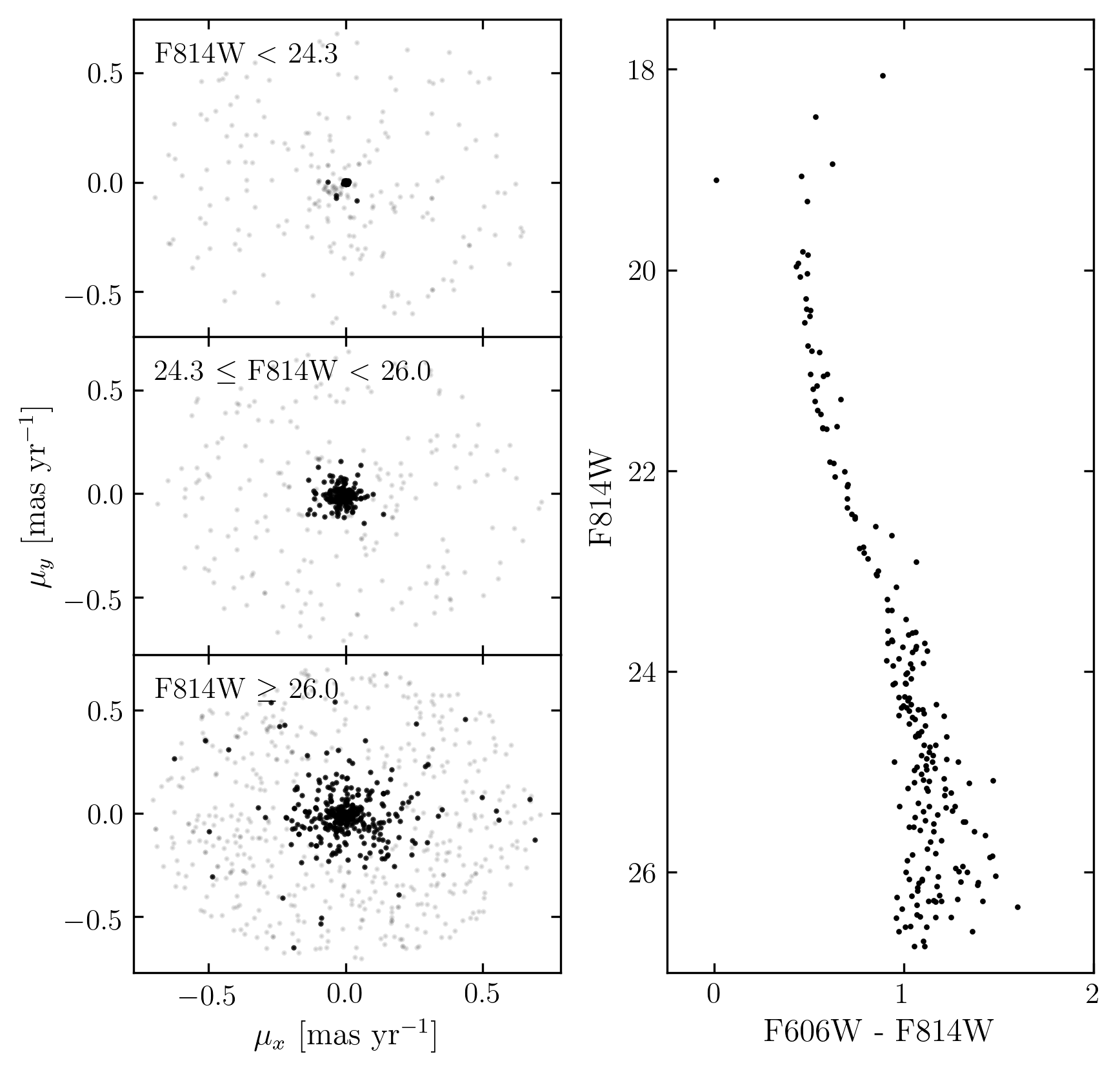}
    \caption{Selection cut based on the relative proper motions for the HST catalogue. \textit{Left:} relative proper motion diagrams of stars in three F814W magnitude bins, with stars selected in black and rejected objects in grey, based solely on proper motion. \textit{Right:} CMD after removing contaminants, based on the proper motion selection and on two parameters quantifying the quality of the ePSF fit.}
    \label{fig:selection_cuts}
\end{figure}

This proper-motion selection is a continuous function of F814W magnitude, reflecting the increasing uncertainties for fainter stars. Bright stars, for which the astrometry is more precise, are required to have proper motions relative to Dra\,II close to zero to be selected as members, while we apply more permissive cuts to fainter stars. This allows us to retain a larger fraction of faint stars, at the cost of admitting some faint contaminants. This magnitude-dependent selection is shown in the left panels of Fig.~\ref{fig:selection_cuts}, where three representative magnitude bins illustrate how the selection radius widens toward fainter magnitudes. Within each bin, the proper-motion selection radius varies, so stars with the same proper motion may be accepted or rejected depending on their magnitude. This effect is particularly clear in the faintest bin (bottom left panel). While we calculated only relative proper motion of the stars---which serves our purpose of isolating contaminants based on their relative velocities---\citet{Warfield:2025} used constructed reference frames to obtain the absolute proper motion of stars in Dra\,II. For the JWST data, however, only a single epoch is available, so relative proper motions cannot be measured and the catalogue may therefore be more contaminated.

Following the proper-motion selection (HST only), we apply quality cuts to both HST and JWST catalogues based on the two parameters that quantify the quality of the ePSF fit to each detection. This allows to reject non-stellar objects and poorly measured detections. The cleaned colour-magnitude diagram (CMD) after all selection criteria is shown in the right panel of Fig.~\ref{fig:selection_cuts}. This process not only allows us to remove contaminants, but also to identify likely HST field stars---objects passing quality cuts but failing proper-motion selection. These criteria ensures that the objects flagged as field stars are likely stars (according to the quality parameters) but not members of Dra\,II (based on their proper motion). As we do not have proper motion for stars in the JWST catalogue, we cannot confidently identify field stars in the JWST data. Subsequent analyses employ either the HST, JWST, or combined catalogue as appropriate for each scientific goal.

\section{Results}
\label{sec:results}
\subsection{CMD analysis}
\begin{figure}
	\includegraphics[width=\columnwidth]{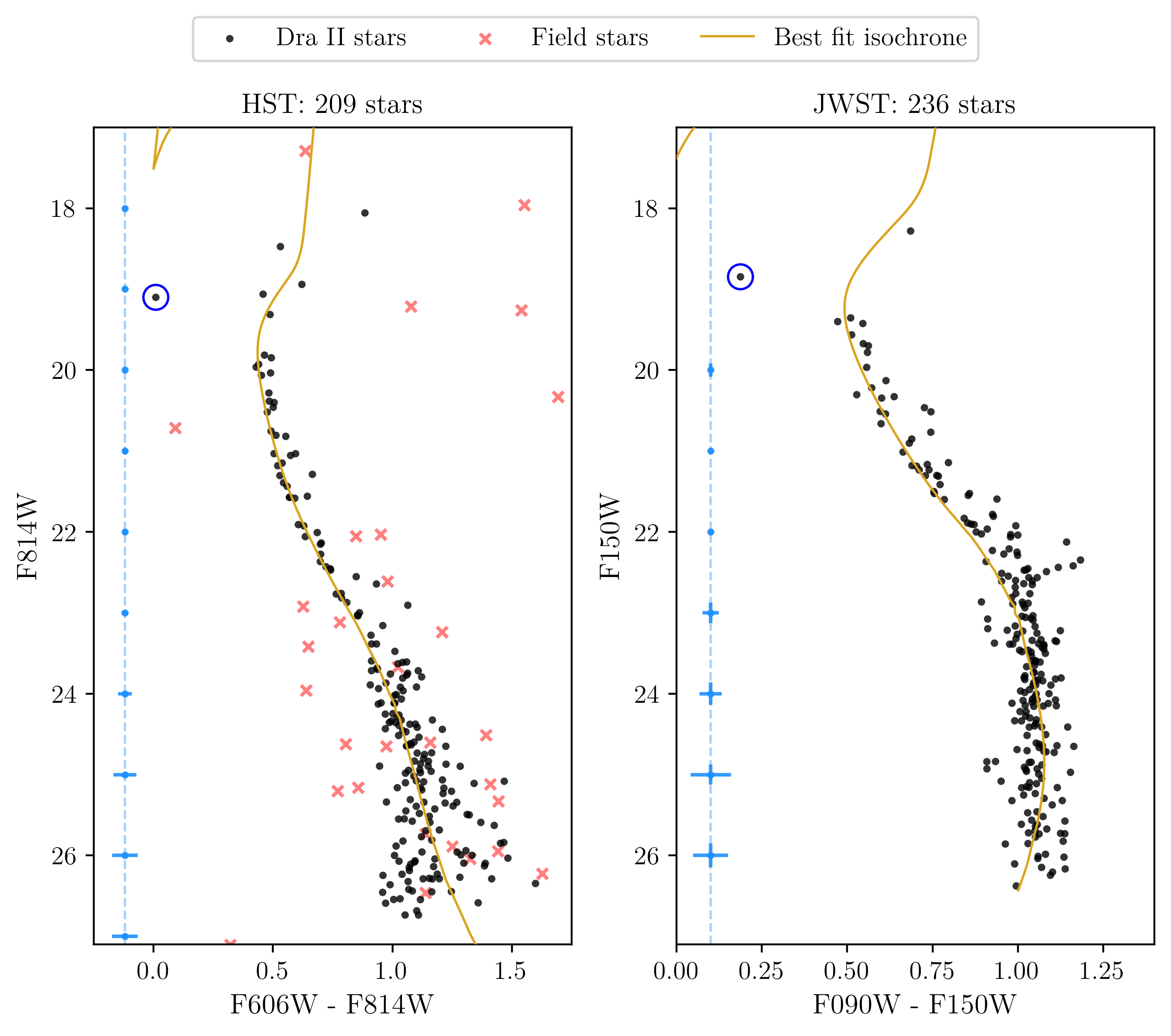}
    \caption{\textit{Left:} HST CMD of Dra\,II with the magnitudes in the filters F606W and F814W, with the best-fitting BaSTI isochrone in orange (d=$19.2\pm0.9$\,kpc, $\feh_{CMD} = -2.5\pm0.2$\,dex, age of $13.5_{-1.5}^{+1.7}$\,Gyr and $\afe$=0.4\,dex). The uncertainties are shown on the left as a function of the F606W magnitude. The completeness is $50\%$ at F814W = 26.5. The field stars are the red crosses. The potential blue straggler star is circled at F606W - F814W $=0.01$ and F814W $=19.10$. \textit{Right:} JWST CMD of Dra\,II with the magnitudes in the filters F090W and F150W. Similarly, the best fit isochrone, the uncertainties and the potential BSS are shown in this CMD.}
    \label{fig:CMD_isochrone}
\end{figure}

The HST-based and JWST-based CMDs of Dra\,II members are displayed in Fig.~\ref{fig:CMD_isochrone}, where the high quality of the optical and infrared photometry can be appreciated. We also show the uncertainty (blue points with error bars) as a function of the F814W and F150W magnitudes. With artificial star tests we determined the 50 percent detection threshold at $F814W = 26.5$\,mag (corresponding to the absolute magnitude $M_{F814W} = 10.1$\,mag). Completeness drops rapidly at fainter magnitudes, as the stars become increasingly difficult to detect and to measure with sufficient precision to pass our proper-motion and quality-selection cuts. In the HST panel, the field stars are indicated as red crosses. The single star circled in blue is a potential blue straggler star (BSS). Both CMDs highlight a well-defined main-sequence (MS) with a faint end well populated, owing to the superb photometry. Due to its extremely low stellar content, Dra\,II features no horizontal branch ($F814W \sim 17.5$) and only a few main-sequence turn-off stars.

We fit BaSTI isochrones \citep{Hidalgo:2018} to the MS of Dra\,II using the most populated CMD with the lowest uncertainties, namely the F814W vs F606W$-$F814W CMD. To derive the best-fit isochrone, we represent each Dra\,II star as an ellipse based on its uncertainties in colour and magnitude. We then compute the total distance from all stars to a given isochrone. Repeating this calculation over a grid of isochrones spanning age, distance, $\feh$, and $\afe$, we identify the best-fitting isochrone as the one that minimises this total distance. Nonetheless, the fit parameters of age, metallicity, $\alpha$-enhancement, and distance show some degeneracy given the small number of MS turn-off stars available. In particular, we investigate three different ages that fit the turn-off region well: 12.5, 13.5 and 14.5\,Gyr, keeping the other parameters unchanged. We compare the distribution of stellar masses as well as the mass function parameters obtained from each of the three isochrones and conclude that the inferred stellar masses do not vary significantly within this age range. Despite the small number of stars in the vicinity of the MS turn-off, there is no evidence for a possible age range among Dra\,II members. Similarly, we investigate isochrones with $\afe = 0.0$ and $\afe = +0.4$\ dex. Previous studies of the stellar population of Dra\,II consistently find an $\alpha$-enhanced composition, but do not report any internal distribution or trend of $\afe$ with $\feh$. Moreover, \citet{Vargas:2013} showed that UFDs generally host $\alpha$-enhanced stellar populations at low metallicity. However, while some UFDs exhibit a decline of $\afe$ with increasing $\feh$, others show little to no significant evolution and remain consistently $\alpha$-enhanced across their full metallicity range. Given the absence of a universal $\afe$--$\feh$ relation among ultra-faint dwarf galaxies, and the absence of observational constraints on a possible spread in the stellar population of Dra\,II, we therefore adopt a single, constant positive value of $\afe=+0.4$. This value is also consistent with other Milky Way dwarf galaxies and disrupted systems, such as \textit{l'itoi} \citep[$\afe = 0.38$,][]{Naidu:2022}, \textit{Thamnos} \citep[$\afe = 0.29$,][]{Naidu:2022}, and \textit{Ursa Minor} \citep[$\afe = 0.4$,][]{Sato:2025}. A discussion about this choice and the impact on the results is presented in Appendix~\ref{app:alpha_tests}. We conclude that the best-fit parameters for Dra\,II are a distance of $19.2\pm0.9$\,kpc, a mean metallicity of $\feh_{CMD} = -2.5\pm0.2$\,dex, an age of $13.5_{-1.5}^{+1.7}$\,Gyr and an $\alpha$-enhancement of $\afe$=0.4\,dex, as illustrated in Fig.~\ref{fig:CMD_isochrone}. These best-fit parameters differ slightly from those reported in previous studies. However, the isochrone fitting is intrinsically limited by the small number of confirmed Dra\,II members, resulting in a poorly populated MS turn-off and the absence of a horizontal branch. Consequently, the derived age, metallicity, $\alpha$-enhancement, and distance remain subject to degeneracies.

We note a slight overabundance of ``stars'' on the red side of the best fitting isochrones. These objects are unresolved binary stars, which we will discuss in the next section.

The chemical properties of Dra\,II provide a way to constrain its stellar mass and evolutionary history. The relation presented by \citet{Naidu:2022} correlates the $\alpha$-enhancement with the quantity $\Delta\feh$, which measures how metal-poor (or metal-rich) a system is relative to expectations from its stellar mass at the present epoch. From this quantity, we can estimate the stellar mass required to reproduce the observed chemical enrichment. Assuming a constant $\afe = +0.4$, we derive $\Delta\feh = (\afe - 0.08)/0.43 \approx 0.74$. Using Eq.\,(5) of \citet{Naidu:2022},
\[
\Delta \feh = -1.66 + 0.30 \log\left(\frac{M_*^{\mathrm{obs}}}{10^6\,M_\odot}\right) - \feh^{\mathrm{obs}},
\]
and adopting $\feh^{\mathrm{obs}} \approx -2.5$, we infer a stellar mass of ${M_* \sim 1.3\times10^5\,M_\odot}$. However, \citet{Fu:2023} declare that the mass--metallicity relation breaks down in the regime of ultra-faint systems, where galaxies with $\log(M_*/M_\odot)\lesssim5$ appear to converge toward a nearly constant metallicity around $\feh\sim-2.6$. Since metal-poor systems are also generally $\alpha$-enhanced, this suggests that metallicity and $\alpha$-enhancement alone cannot robustly constrain the stellar mass of Dra\,II. Consequently, the value derived from the relation of \citet{Naidu:2022} should be interpreted cautiously and may primarily reflect the chemical enrichment history of the system rather than its true stellar mass.

This stellar mass derived from chemical composition is substantially larger than the present-day stellar mass of Dra\,II derived from its luminosity. Assuming a purely baryonic system, \citet{Longeard:2018} estimated a stellar mass of only $\sim360\,M_\odot$. The discrepancy suggests that Dra\,II was originally significantly more massive and subsequently lost a large fraction of its stars through tidal stripping. Such a scenario is plausible given the orbit derived by \citet{Longeard:2018}, with a pericenter of $\sim21$\,kpc and an apocenter of $\sim150$\,kpc. 

The present-day stellar mass is also much smaller than the dynamical mass estimated from the velocity dispersion and half-light radius, $M_{1/2}\approx3\times10^6\,M_\odot$ \citep{Martin:2016}. Unlike stellar mass, the dynamical mass traces the total gravitating mass and therefore includes any dark matter halo. However, the estimate of the dynamical mass assumes virial equilibrium, which may not hold if Dra\,II is undergoing tidal disruption. Nevertheless, even accounting for these uncertainties, the present-day stellar mass remains orders of magnitude smaller than the present-day dynamical mass inferred from the velocity dispersion. Unless Dra\,II's velocity dispersion is substantially inflated by tidal effects, this large discrepancy strongly suggests that Dra\,II is significantly dark-matter dominated.

\subsection{Unresolved binary stars}

We investigate unresolved binaries, i.e. systems in which the stars are so close that their light is blended and the system appears as a single point source. Following \citet{Milone:2012} and \citet{Cordoni:2018}, we define the range of mass ratios and magnitudes over which unresolved binaries can be reliably identified in our data.

In the CMD, the position of an unresolved binary system depends on its primary mass $M_1$ and mass ratio $q = M_2/M_1$. Because the fluxes of both components add, the combined system appears brighter than a single star of the same primary mass: \[m_{bin} = m_1 - 2.5\log\left(1+\frac{F_2}{F_1}\right)\] Equal-mass binaries ($q=1$) therefore form a sequence almost parallel to the MS, located about $0.75$\,mag brighter. Systems with $q < 1$ are still brighter and redder than the single-star MS, but they lie below the equal-mass binary sequence. This motivates the investigation of unresolved main-sequence binary stars in the F814W vs F606W$-$F814W CMD, which provides the highest photometric precision and is the least sensitive to intrinsic metallicity variations. Hence, any significant deviation from the MS can be reasonably attributed to an unresolved binary star. We restrict the analysis to the MS stars with $20.8 \leq m_{F814W} \leq 23.8$, corresponding to primary masses in the range $0.43 \leq M_1 / M_\odot \leq 0.71$. This avoids the MS turn-off region, where metallicity effects are stronger, and the faint lower MS, where photometric uncertainties become large.

\begin{figure*}
    \centering
    \includegraphics[width=1\linewidth]{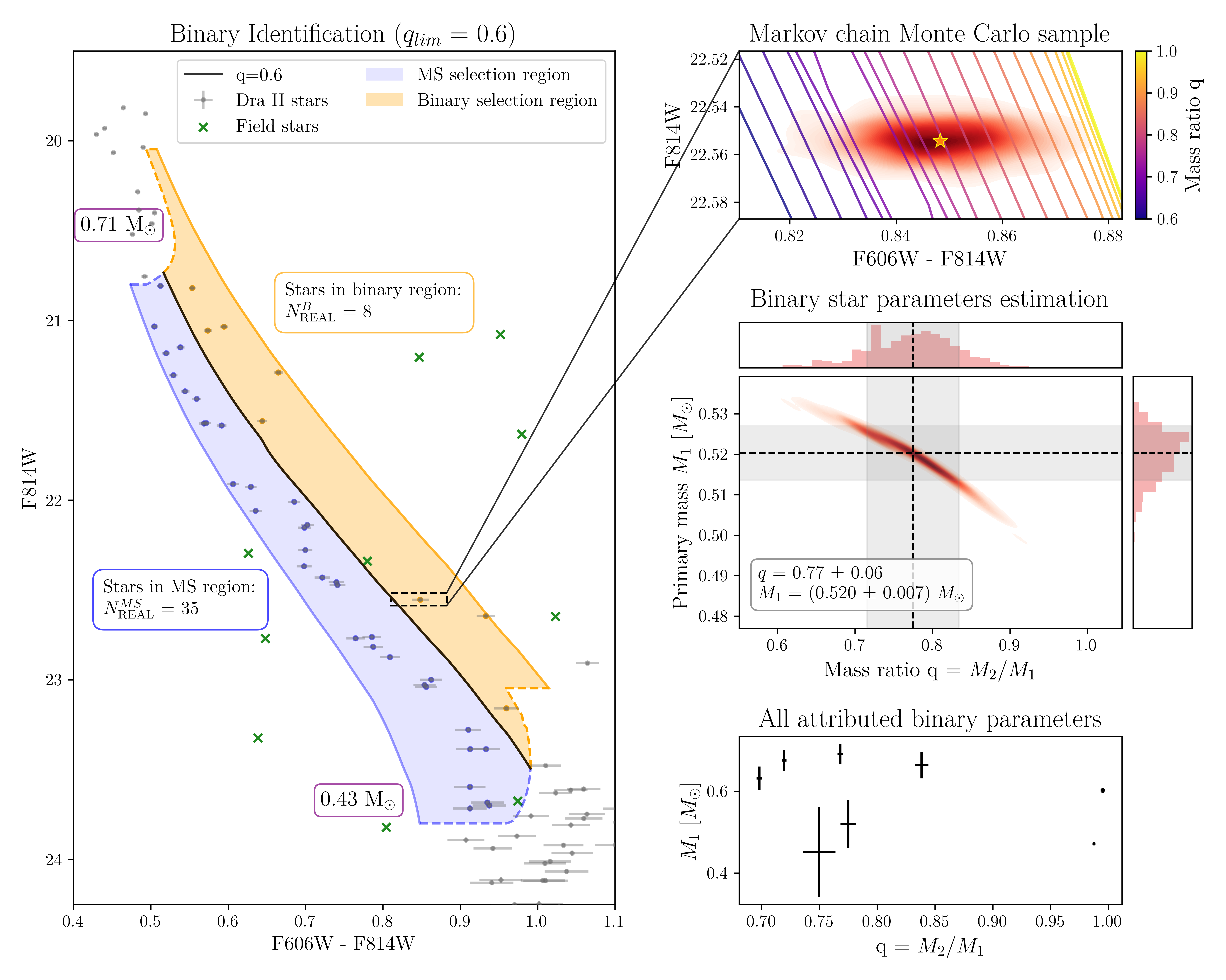}
    \caption{\textit{Left:} Definition of the regions to count MS stars and unresolved binaries for $q_{\rm lim} = 0.6$. The main-sequence region (blue) is defined as the area between the best-fit isochrone shifted by $3\sigma_{F606W-F814W}$ to the left and the binary sequence corresponding to $q=q_{\rm lim}$. The binary region (orange) is defined as the area between the $q=q_{\rm lim}$ binary sequence and the equal-mass binary sequence shifted by $3\sigma_{F606W-F814W}$ to the right. Field stars are indicated as green crosses. Each of the 8 Dra\,II stars falling in the binary region is then modelled by a Monte Carlo simulation accounting for photometric uncertainties; an example of one identified binary star and its associated Monte Carlo simulation is shown in the \textit{top right} panel. We obtain a distribution of $q$ and $M_1$ using a grid of fiducial binary tracks---some are shown in the same panel for different values of $q$. The final parameters $M_1$ and $q$ inferred for each binary, along with their uncertainties, are derived from the median and standard deviation of the Monte Carlo distributions, as illustrated in the \textit{middle right} panel. The binary parameters of all 8 binary systems, along with their uncertainties, are represented in the \textit{bottom right} panel.}
    \label{fig:binaries_all}
\end{figure*}

Since binaries with smaller mass ratios lie closer to the single-star main sequence and are harder to detect, we express the inferred binary fraction as a function of $q_{\rm lim}$, the minimum mass ratio that can be reliably identified given the photometric uncertainties. We repeat the analysis for $q_{\rm lim} = 0.5$, $0.6$ and $0.7$. For each value of $q_{\rm lim}$, we define two regions within the selected mass range, separated by the fiducial sequence corresponding to $q=q_{\rm lim}$, to count the MS stars and unresolved binaries following \citet{Milone:2012}. These two regions are illustrated in the left panel of Fig.~\ref{fig:binaries_all} for $q_{\rm lim} = 0.6$. We count the stars in the MS and binary region to obtain $N_{\rm REAL}^{MS}$ and $N_{\rm REAL}^{B}$, respectively. The same procedure is then applied for $q_{\rm lim} = 0.5$ and $q_{\rm lim} = 0.7$.

For each unresolved binary candidate, we estimate the primary mass $M_1$ and the mass ratio $q$ by comparing its observed position in the CMD to a grid of theoretical binary sequences. To account for photometric uncertainties, we propagate the errors using Monte Carlo simulations: we draw realisations of the colour and magnitude from normal distributions centered on the observed values, with standard dispersions equal to the photometric errors. An example of the resulting distribution in the CMD is shown in the top-right panel of Fig.~\ref{fig:binaries_all}. For each draw, we assign a primary mass and a mass ratio by interpolating within a grid of binary tracks (some are shown in the same panel). The final estimates of $M_1$ and $q$, together with their uncertainties, are taken as the median and standard deviation of the resulting Monte Carlo distributions (see the middle-right panel for the joint distribution of $M_1$ versus $q$ for one of the binary stars). We also infer other quantities, such as the F814W and F606W magnitudes of the individual components. The bottom-right panel of Fig.~\ref{fig:binaries_all} summarises the primary mass and mass ratio estimates, with their associated uncertainties, for the eight binaries candidates.

From the identification of unresolved binary stars, we can evaluate the binary fraction and its uncertainty for the three different values of $q_{\rm lim}$. Following Eq.\,1 of \citet{Milone:2012}, we compute \[f_{bin}^{q>q_{\rm lim}} = \frac{N_{\rm REAL}^B - N_{\rm FIELD}^B}{N_{\rm REAL}^{MS + B} - N_{\rm FIELD}^{MS + B}} - \frac{N_{\rm ART}^B}{N_{\rm ART}^{MS + B}},\] where the superscript $B$ refers to the binary region, and $MS+B$ denotes the combined MS and binary regions. As in \citet{Milone:2012} and \citet{Cordoni:2018}, we correct the binary fraction for field contamination and for the presence of blended stars introduced during the data reduction process, quantified using the artificial star catalogue. This correction is small: $N_{\rm ART}^B/N_{\rm ART}^{MS + B} = 0.006, 0.002$ and $0.001$ for $q_{\rm lim} = 0.5, 0.6$ and $0.7$ respectively. To estimate the uncertainty $\sigma_{f_{bin}}$, we generate bootstrap resamples of the CMD. For each sample, we perturb the colour and magnitude of every star by adding a Gaussian noise representative of the photometric uncertainties. We then recompute the binary fraction for each resample and for all three $q_{\rm lim}$ values. The standard deviation of the resulting distribution is adopted as the uncertainty on the binary fraction. 

We apply the same procedure to the JWST catalogue to identify unresolved binaries and measure their properties. Figures for the JWST catalogue are in the Appendix~\ref{app:jwst_figs}. The F150W vs F090W$-$F150W CMD is mostly unaffected by metallicity variations, making it well suited for this purpose. Given the similar spatial resolution of JWST and HST, we assume that field contamination and blending effects are comparable in the two catalogues. This assumption has little impact on the final result: if the number of field stars in the JWST catalogue is under- or overestimated by 30$\%$, the resulting binary fraction changes by only $\pm 0.01$, which remains well within the uncertainty range. We therefore derive the corrected JWST binary fractions for the three $q_{\rm lim}$ values. The final HST and JWST binary fractions are reported in Table~\ref{tab:binary_fraction}.

\begin{table}
\small
\centering
\caption{Dra\,II binary fractions for three values of $q_{\rm lim}$ derived from HST and JWST photometry.}
\begin{tabular}{c c c c}
\hline\hline
$ q_{\rm lim} $ & $ N_{\mathrm{\rm REAL}}^{B} $ & $ N_{\mathrm{\rm REAL}}^{MS + B} $ & $ f_{\mathrm{bin}}^{q>q_{\rm lim}}  \pm \sigma_{f_{bin}}^{q>q_{\rm lim}}$\\
\hline
\multicolumn{4}{c}{HST}\\\hline

0.5 & 10 & 43 & $0.21 \pm 0.07$ \\
0.6 & 8 & 43  & $0.17 \pm 0.06$\\
0.7 & 8 & 43  & $0.17 \pm 0.06$\\
\hline

\multicolumn{4}{c}{JWST}\\\hline
0.5 & 12 & 41 & $0.28 \pm 0.07$\\
0.6 & 11 & 41  & $0.25 \pm 0.06$\\
0.7 & 9 & 41  & $0.20 \pm 0.06$\\
\hline

\end{tabular}
\label{tab:binary_fraction}
\end{table}

Following \citet{Milone:2012}, we adopt a probabilistic approach to infer the total binary fraction $f_{\rm bin}^{\rm total}$ from the measured binary fractions $f_{\rm bin}^{q>q_{\rm lim}}$ at different mass‑ratio limits $q_{\rm lim}$. Since the true mass‑ratio distribution of binaries in Dra\,II is unknown, and the number of identified binaries remains limited, we choose a conservative approach and infer the total binary fraction using a flat distribution in $q \in [0,1]$, which yields $f_{\rm bin}^{\rm total} = f_{\rm bin}^{q>q_{\rm lim}}/(1 - q_{\rm lim}) = 0.5 \pm 0.1$. This value can vary between $0.3$--$0.7$ depending on the distribution profile chosen (favouring equal-mass binaries or favouring low mass ratios). This result will be discussed in Section~\ref{sec:discussion}.


\subsection{Metallicity distribution of bright Dra II stars}
\label{sec:cahk}

Metallicity, along with other stellar parameters such as age and element abundance ratios, affects photometric magnitudes and colours depending on the filter's wavelength coverage. By selecting appropriate filter combinations to define pseudo-colours, the effects of intrinsic variations in these stellar properties can be enhanced in colour–colour diagrams constructed for different regions of the CMD (MS, red giant branch, asymptotic giant branch, etc.). One useful colour-colour diagram  is the CaHK diagram, inspired by the Pristine survey \citep{Starkenburg:2017}. Narrow-band photometry around the H and K lines of the singly ionised calcium (Ca\,II), with wavelengths respectively at $396.8$ and $393.4$\ nm, provides a good estimate of the metallicity of the star. Indeed, metal-poor stars show weaker Ca H\&K absorption features compared to more metal-rich stars of the same temperature. Moreover, the strength of these absorption lines depends only weakly on other stellar parameters like surface gravity: using other filters to extract the temperature of a star, one can then infer the metallicity from CaHK photometry. The HST F395N filter is a good analogue to the narrow-band CaHK filter, as it covers the absorption lines Ca H\&K \citep{Fu:2023}. In particular, \cite{Fu:2022} demonstrated that metallicities derived from F395N align closely with spectroscopic calcium triplet calibration \citep{Li:2017}. 

\begin{figure}
    \centering
    \includegraphics[width=1\linewidth]{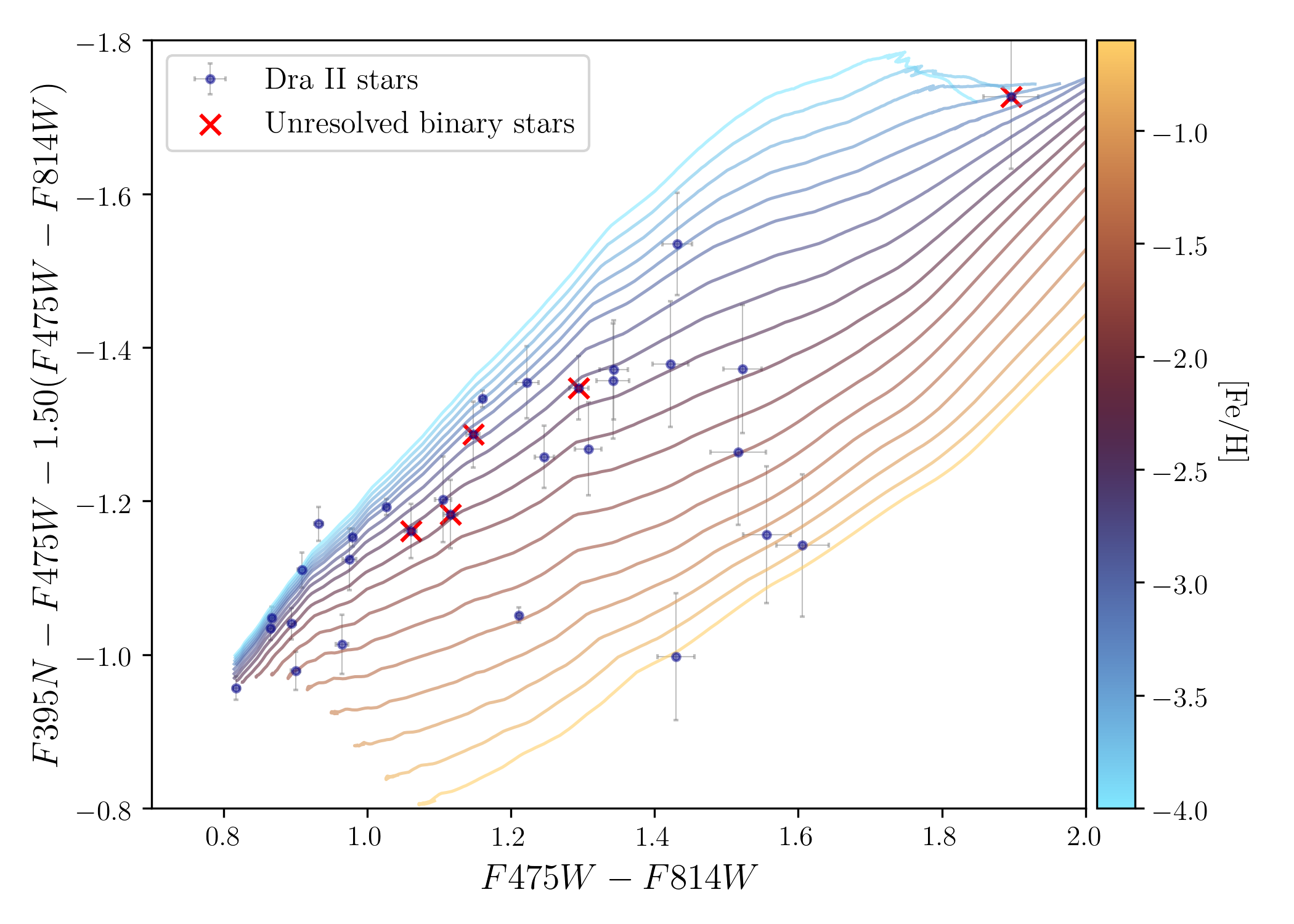}
    \caption{CaHK colour-colour diagram inspired by the Pristine survey \citep{Starkenburg:2017}. Dra\,II stars are the blue points, the identified unresolved binaries are the red crosses. The faintest star has $\mathrm{F814W} = 23$\ mag. A grid of BaSTI isochrones is overplotted, for different metallicities.}
    \label{fig:cahk}
\end{figure}

We construct the HST-based CaHK diagram using ${\mathrm{F475W}-\mathrm{F814W}}$ on the $x$-axis (colour sensitive to temperature) and ${\mathrm{F395N}-\mathrm{F475W}-1.5\times(\mathrm{F475W}-\mathrm{F814W}}$) on the $y$-axis \citep[pseudo-colour sensitive to metallicity;][]{Fu:2023}. This colour–colour diagram (Fig.~\ref{fig:cahk}) effectively traces metallicity spreads among MS stars. Because the diagram requires photometry in the narrow F395N passband, this analysis is restricted to brighter stars (${\mathrm{F814W} = 18.5-23.0\,\mathrm{mag}}$). The data points include photometric uncertainties, propagated from artificial star tests. We overplot a grid of isochrones having $\afe = +0.4$\,dex, an age of ${13.5\ \mathrm{Gyr}}$, and metallicities ranging from $\feh = -4.0$ to $-0.5$ in steps of $0.1$\ dex, truncated to their MS portions. We correct the isochrones for the photometric bias obtained from the artificial star tests, defined as the median difference between input and output magnitudes, following the method of \citet{Fu:2023}. The bias is negligible in F814W and F606W, and is dominated by the F395N filter. Accounting for this bias systematically increases the inferred metallicity. Some isochrones are shown in Fig.~\ref{fig:cahk}. Per‑star metallicities are then derived by matching individual stars to this grid of isochrones: for each star, we first generate 1000 realisations by perturbing the three magnitudes F395N, F475W, and F814W according to their individual photometric errors, recomputing the colours in each case. We then calculate the minimum distance to every isochrone track, in CaHK space, and assign the metallicity $\feh$ of the nearest track. The assigned metallicities from these realisations are then used to compute a median value and a standard deviation: the median provides the best‑estimate per‑star metallicity, while the standard deviation yields the per‑star uncertainty on $\feh$. This method accounts both for measurements errors and the intrinsic diagram sensitivity to metallicity---at bluer F475W$-$F814W, the strength of Ca II H and K absorption lines decreases rapidly, and the isochrone tracks are closer to each other, leading to larger uncertainties on the per-star $\feh$.

\begin{figure}
    \centering
    \includegraphics[width=1\linewidth]{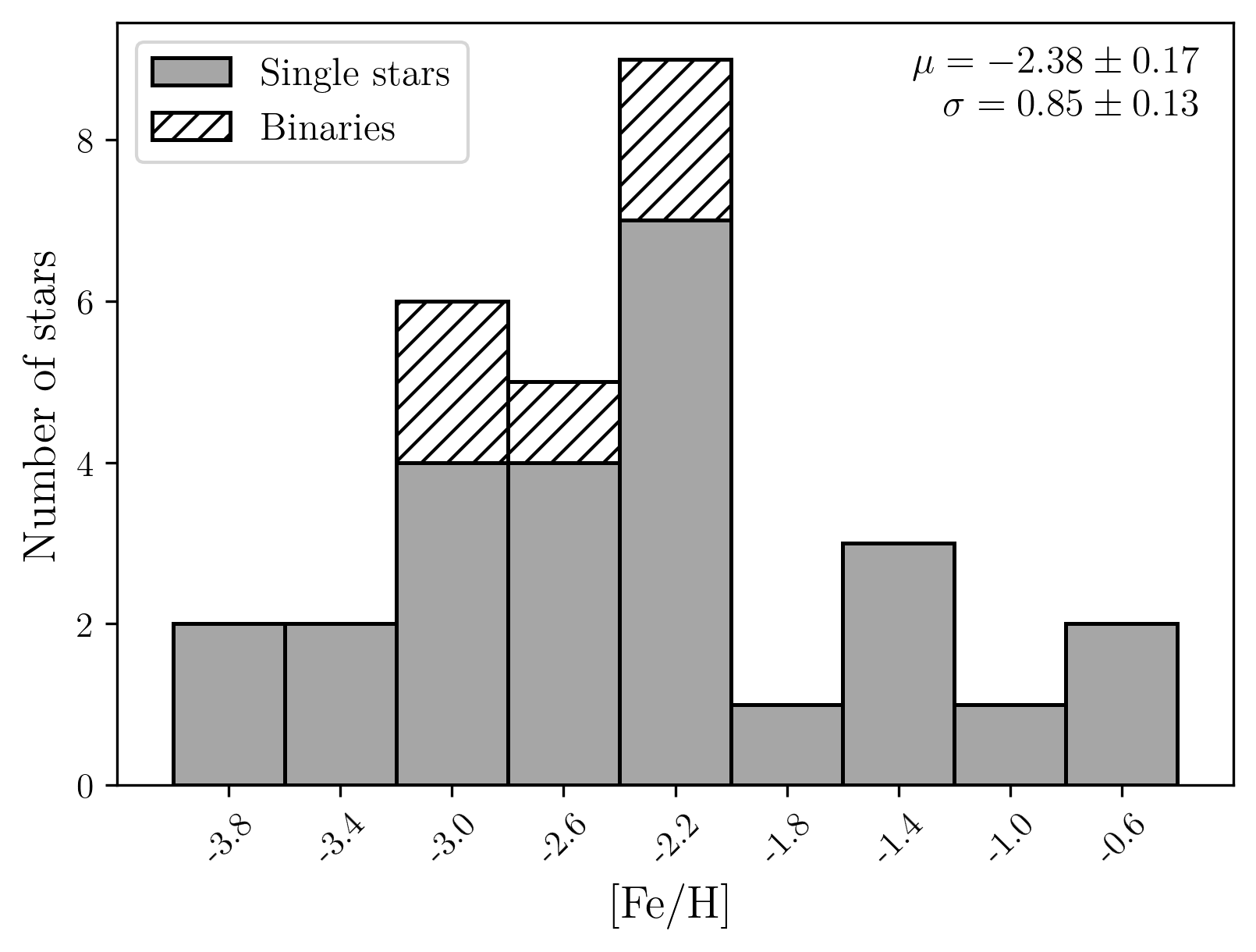}
    \caption{Metallicity distribution of the bright stars derived from the CaHK diagram, shown as the number of stars per metallicity bin. We characterise the data excluding binaries with $\feh_{CaHK} = -2.38 \pm 0.17$\,dex and $\sigma_{\feh} = 0.85 \pm 0.13$\,dex.}
    \label{fig:met_cahk}
\end{figure}

The final metallicity distribution is shown in Fig.~\ref{fig:met_cahk}, and stars that were previously identified as binaries are also illustrated in this figure. We then model the unbinned $\feh$ distribution, excluding binary stars, following the method adopted by \citet{Bissonette:2026}. We compare a single Gaussian to the distribution and determine its best-fit parameters $\mu$ and $\sigma$ by minimising the negative log-likelihood: \[\ln \mathcal{L}(\mu,\sigma,w\}) =\sum_{i=1}^{N}\ln \left(w \,\mathcal{N}\!\left([\mathrm{Fe/H}]_i \mid \mu,\,\sqrt{\sigma^2 + \sigma_i^2}\right)\right)\] where $\mathcal{N}\!\left([\mathrm{Fe/H}]_i \mid \mu,\,\sigma\right)$ is the Gaussian probability density for star $i$. Measurement uncertainties are incorporated by convolving each Gaussian with the individual metallicity error $\sigma_i$ of the star. The model selection is then performed using the Akaike Information Criterion (AIC), $\mathrm{AIC} = 2k-2\ln \mathcal{L}$, where $k$ is the number of free parameters \citep{Akaike:1974}. The model with the lowest AIC value is statistically preferred, balancing goodness-of-fit against model complexity. The uncertainties on the best-fit parameters are estimated analytically from the fitted dispersion and the sample size, namely $\delta\mu = \sigma/\sqrt{N}$ and $\delta\sigma = \sigma/\sqrt{2N}$, where $N$ is the number of stars in the fit. As expected, Dra\,II hosts a metal-poor stellar population, with $\feh_{CaHK} = -2.38 \pm 0.17$\,dex. We also find a significant metallicity spread among the brighter stars, $\sigma_{\feh} = 0.85 \pm 0.13$\,dex. This result will be discussed in Section~\ref{sec:discussion}.

\subsection{Metallicity distribution of faint Dra II stars }
\label{sec:dynesty}

\begin{figure}
    \centering
    \includegraphics[width=1\linewidth]{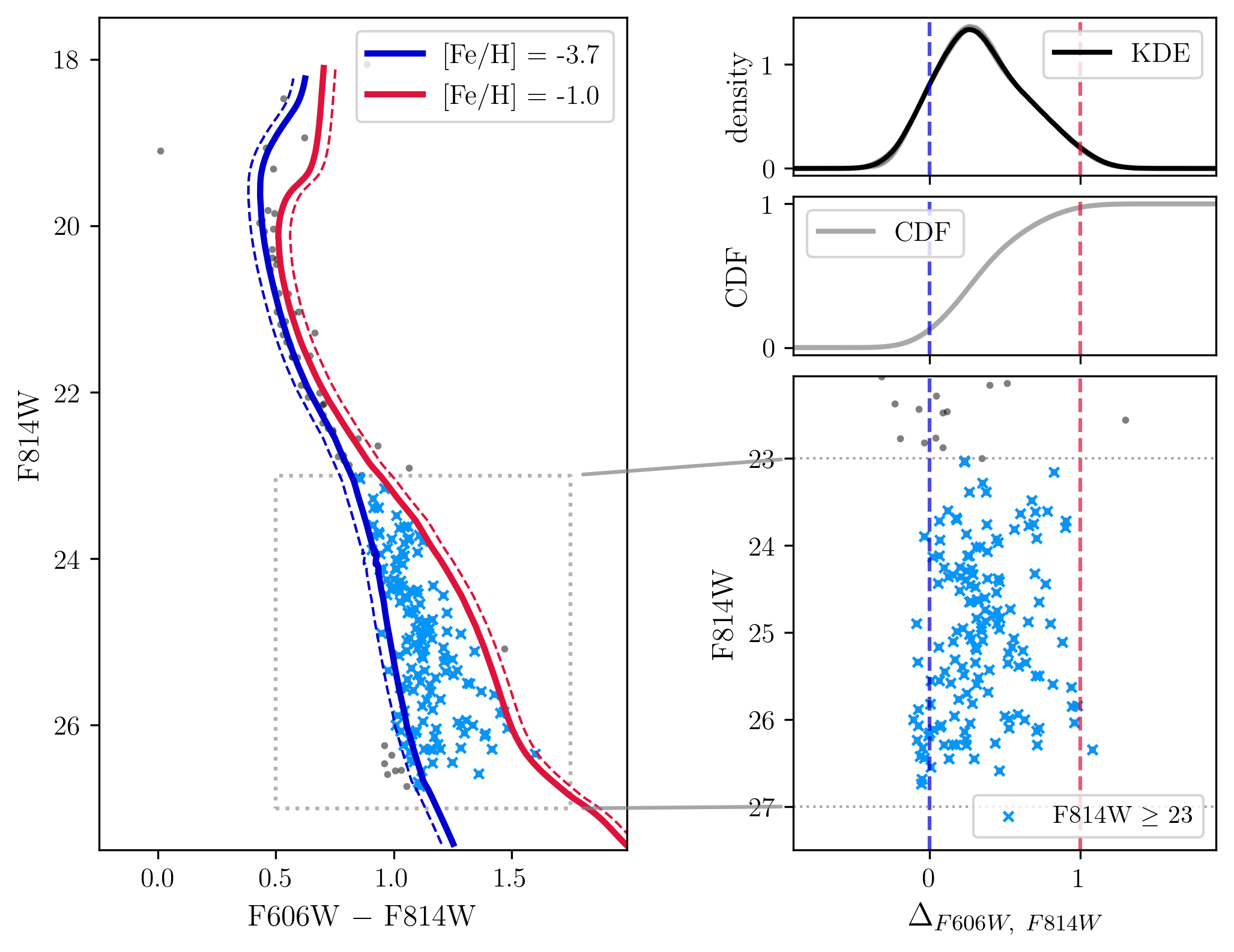}
    \caption{\textit{Left:} Observed CMD of Dra\,II with a metal-poor (blue) and metal-rich (red) isochrone. Both isochrones are $\alpha$-enhanced ($\afe = +0.4$\ dex); the dashed coloured lines correspond to the isochrones shifted by $0.05$\ dex in colour to include in the analysis stars that are just outside the boundaries defined by the isochrones. \textit{Right:} Kernel density estimate (KDE), cumulative distribution function (CDF) and verticalized CMD of the faint stars of Dra\,II.}
    \label{fig:vert_cmd}
\end{figure}

One limitation of the CaHK analysis is that it requires photometry in both the F395N and F475W passbands. These filters do not reach the same depth as F814W or F606W, and the narrow F395N band provides reliable magnitudes only for the brightest stars ($\mathrm{F814W} \leq 23.0$\ mag). To exploit the deeper photometry, we adopt an alternative method to infer the metallicity distribution of stars fainter than $\mathrm{F814W} = 23$ using the more populated F814W vs F606W$-$F814W CMD. As illustrated by the blue and red isochrones in Fig.~\ref{fig:vert_cmd} (age of $13.5$\,Gyr, $\afe = +0.4$, $\feh = -3.7$ and $-1.0$ respectively), metallicity variations significantly impact stars fainter than $\mathrm{F814W} \sim 23$\ mag. We therefore follow an approach similar to that introduced by \citet{Muratore:2025} when they inferred the $\feh$ distribution of the ultra-faint dwarf galaxy Bootes\,I from JWST photometry.

The first step consists of verticalizing the CMD using two isochrones, one metal-rich and one metal-poor, chosen to enclose the bulk of stars fainter than $\mathrm{F814W} \sim 23$. The selected isochrones correspond to $\feh=-3.7$ and $\feh=-1.0$, and are shown in Fig.~\ref{fig:vert_cmd} as blue and red lines, respectively. To verticalize the CMD, we define a pseudo-colour as:
\[
\Delta_{F606W,\ F814W}=\frac{X - X_\mathrm{blue}}{X_\mathrm{red} - X_\mathrm{blue}},
\]
where $X=\mathrm{F606W}-\mathrm{F814W}$, $X_\mathrm{blue}$ is the F606W$-$F814W colour of the blue (metal-poor) isochrone and $X_\mathrm{red}$ is that of the red (metal-rich) isochrone.

\begin{figure}
    \centering
    \includegraphics[width=1\linewidth]{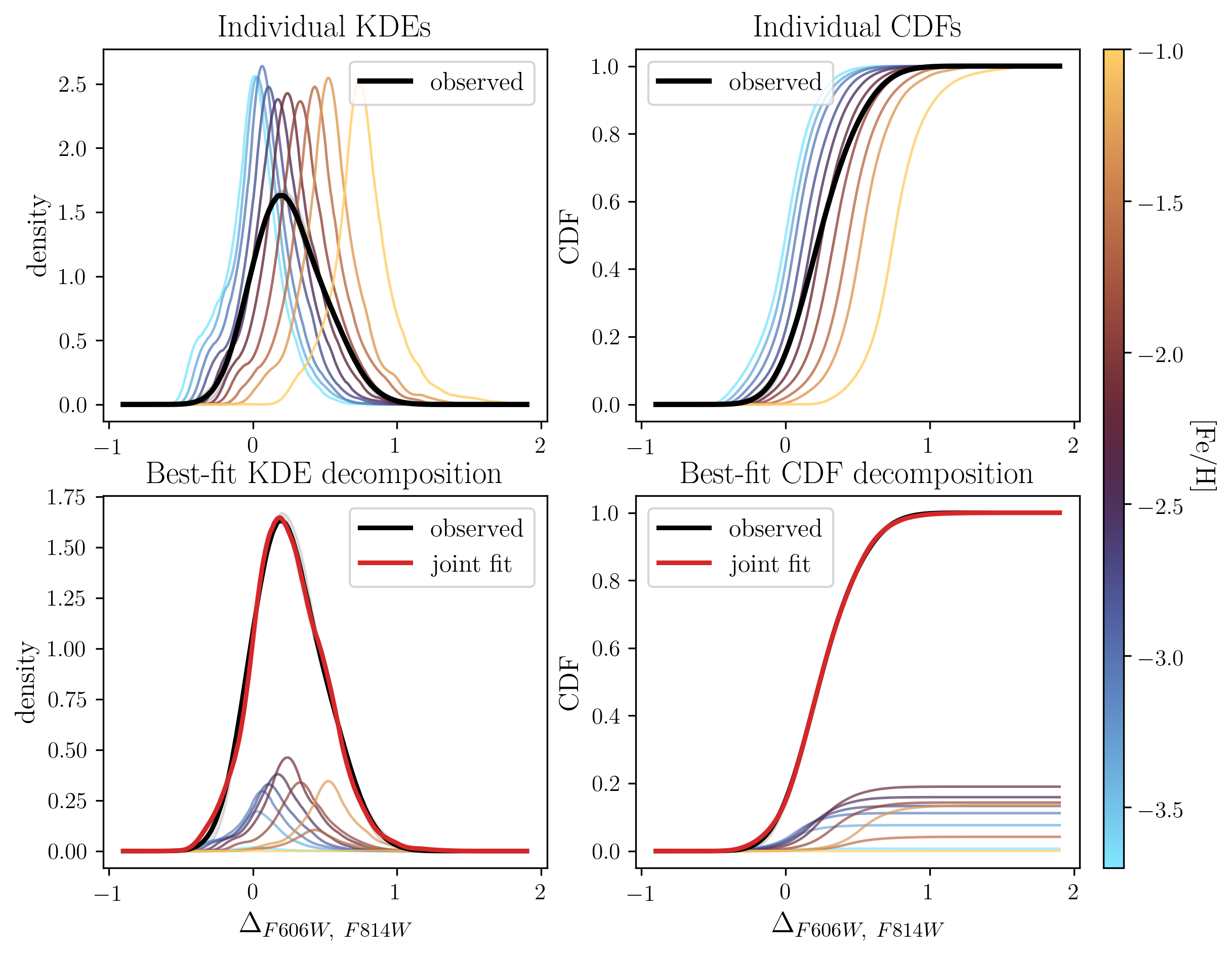}
    \caption{\textit{Top panels:} KDEs and CDFs obtained from the verticalized CMDs of the simulated simple stellar populations with varying metallicities, colour-coded by metallicity. In black are the KDE and the CDF derived from Dra\,II's observed verticalized CMD. \textit{Bottom panels:} Best-fit of the observed KDE and CDF of Dra\,II using a combination of the individual KDEs and CDFs above.}
    \label{fig:kde_cdf}
\end{figure}

We then build the kernel density estimate (KDE) and the cumulative distribution function (CDF) of the pseudo-colour distribution for the faint stars in Dra\,II (right panels of Fig.~\ref{fig:vert_cmd}). The KDE is computed using the \texttt{KernelDensity} implementation from \texttt{sklearn.neighbors}, accounting for statistical uncertainties through bootstrap resampling and for the effect of the bandwidth choice by sampling over a range of bandwidth values. These observed distributions are used as targets to derive the best-fitting metallicity distribution. To this end, we use artificial stars and BaSTI isochrones to simulate simple stellar populations with metallicities ranging from $\feh=-3.7$ to $\feh=-1.0$. We include a simulation of binary stars for each isochrone, with a binary fraction of $0.30$ and a flat mass ratio distribution. Each simulated CMD is then verticalized in the same way as the observed CMD. The simulated CMDs are shown in Appendix~\ref{app:simulated_cmds}. For each simulated and verticalized CMD, we compute the corresponding KDE and CDF, both shown on the first row of Fig.~\ref{fig:kde_cdf} and colour-coded by metallicity. The observed KDE and CDF from Dra\,II's stellar population are shown as black line.

\begin{figure}
    \centering
    \includegraphics[width=1\linewidth]{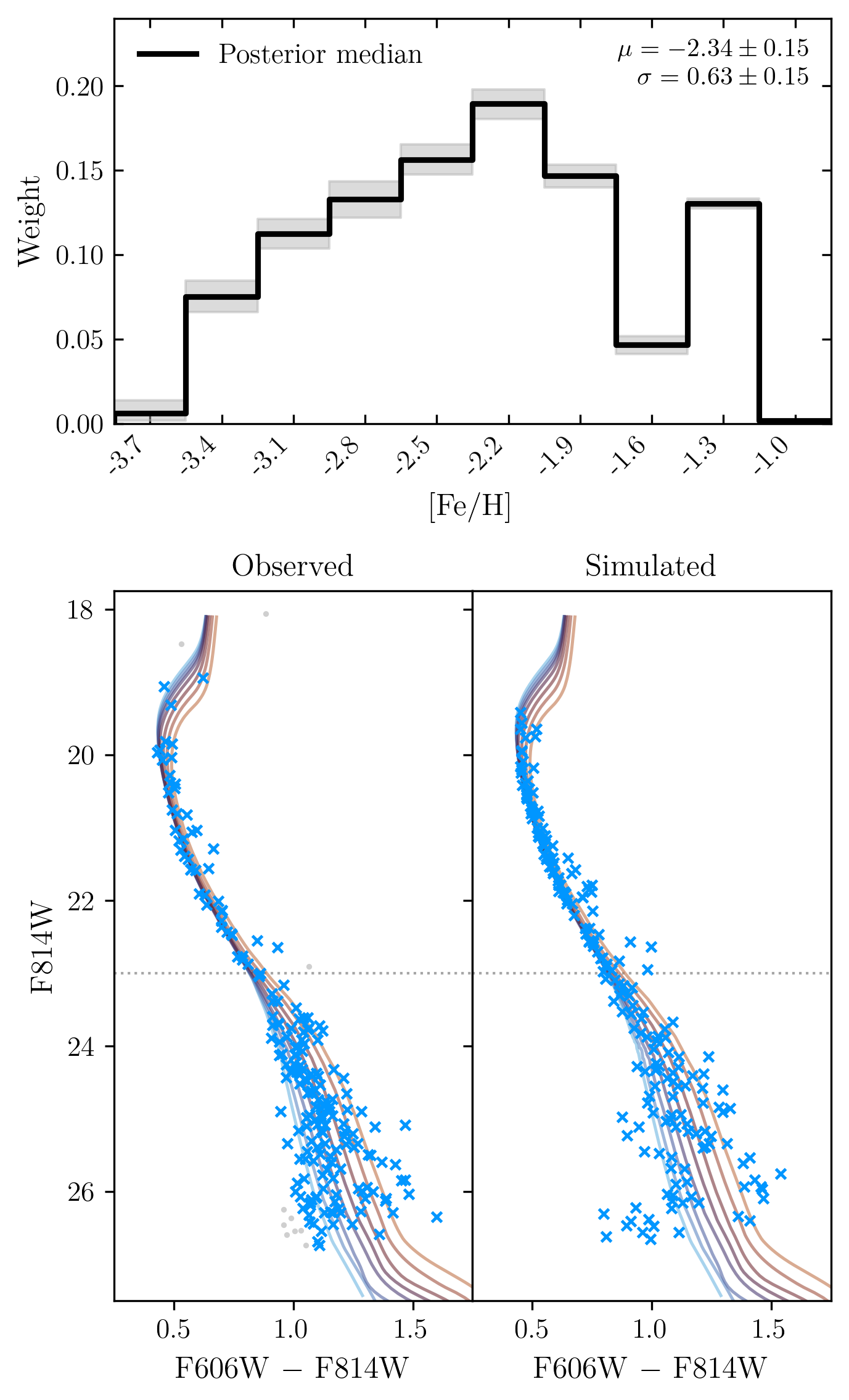}
    \caption{\textit{Top:} Metallicity distribution resulting from the fitting of the observed KDE and CDF, as well as its uncertainties (shaded regions). \textit{Bottom:} Two CMDs of Dra\,II, observed (left) and simulated (right) using the best-fit metallicity distribution function.}
    \label{fig:feh_fit}
\end{figure}

Finally, we use a Bayesian framework to fit both the observed KDE and CDF of Dra\,II as a linear combination of the KDEs and CDFs of the simulated distributions. Each simulated simple stellar population is assigned a weight between 0 and 1, and the final set of weights is normalised to represent the $\feh$ distribution. The optimisation is performed using the \texttt{Dynesty} package \citep{Speagle:2020, Koposov:2025}. The bottom panels of Fig.~\ref{fig:kde_cdf} show the best-fitting models: the total best-fitting KDE and CDF are displayed in red, the observed distributions in black, and the contribution of each metallicity component is shown as coloured lines. The corresponding $\feh$ distribution is presented in the top panel of Fig.~\ref{fig:feh_fit}. This method already accounts for photometric uncertainties, binary stars, and blends introduced by the data reduction; all these effects are illustrated in the shaded region, which represents the uncertainty on the metallicity distribution. We characterise this distribution by its mean and standard deviation, and estimate their uncertainties by drawing 1000 realisations of the sample using Monte Carlo simulations that propagate the weight uncertainties: $\feh_{KDE,CDF} = -2.34\pm0.15$\,dex and $\sigma_{\feh} = 0.63\pm0.15$\,dex. The two bottom panels of Fig.~\ref{fig:feh_fit} show the observed CMD of Dra~\,II (left) and an equivalent simulated CMD constructed from the inferred metallicity distribution, with the same number of stars (right).

We have now obtained a metallicity distribution for bright stars, with
$\feh_{CaHK} = -2.38 \pm 0.17$\,dex and $\sigma_{\feh} = 0.85 \pm 0.13$\,dex, and a metallicity distribution for fainter stars, with $\feh_{KDE,CDF} = -2.34\pm0.15$\,dex and $\sigma_{\feh} = 0.63\pm0.15$\,dex. While the mean metallicity is in good agreement between the two groups, the faint stars seem to have a smaller spread. This result will be discussed in Section~\ref{sec:discussion}.

\subsection{Stellar Mass function of Dra II} 
We assign stellar masses to Dra\,II stars via interpolation using the best-fit isochrone presented in Fig.~\ref{fig:CMD_isochrone} ($\mathrm{d} = 19.2\,\mathrm{kpc}, \feh = -2.5\,\mathrm{dex}, \mathrm{age} = 13.5\,\mathrm{Gyr}, \afe = 0.4\,\mathrm{dex}$). A discussion about the choice of an $\alpha$-enhanced isochrone to assign  masses is presented in Appendix~\ref{app:alpha_tests}. All stars are treated as single stars; the correction to account for the presence of binary stars will be applied in a later stage. We obtain masses ranging from $0.13$ to $0.77\ M_\odot$.

\begin{figure}
    \centering
    \includegraphics[width=1\columnwidth]{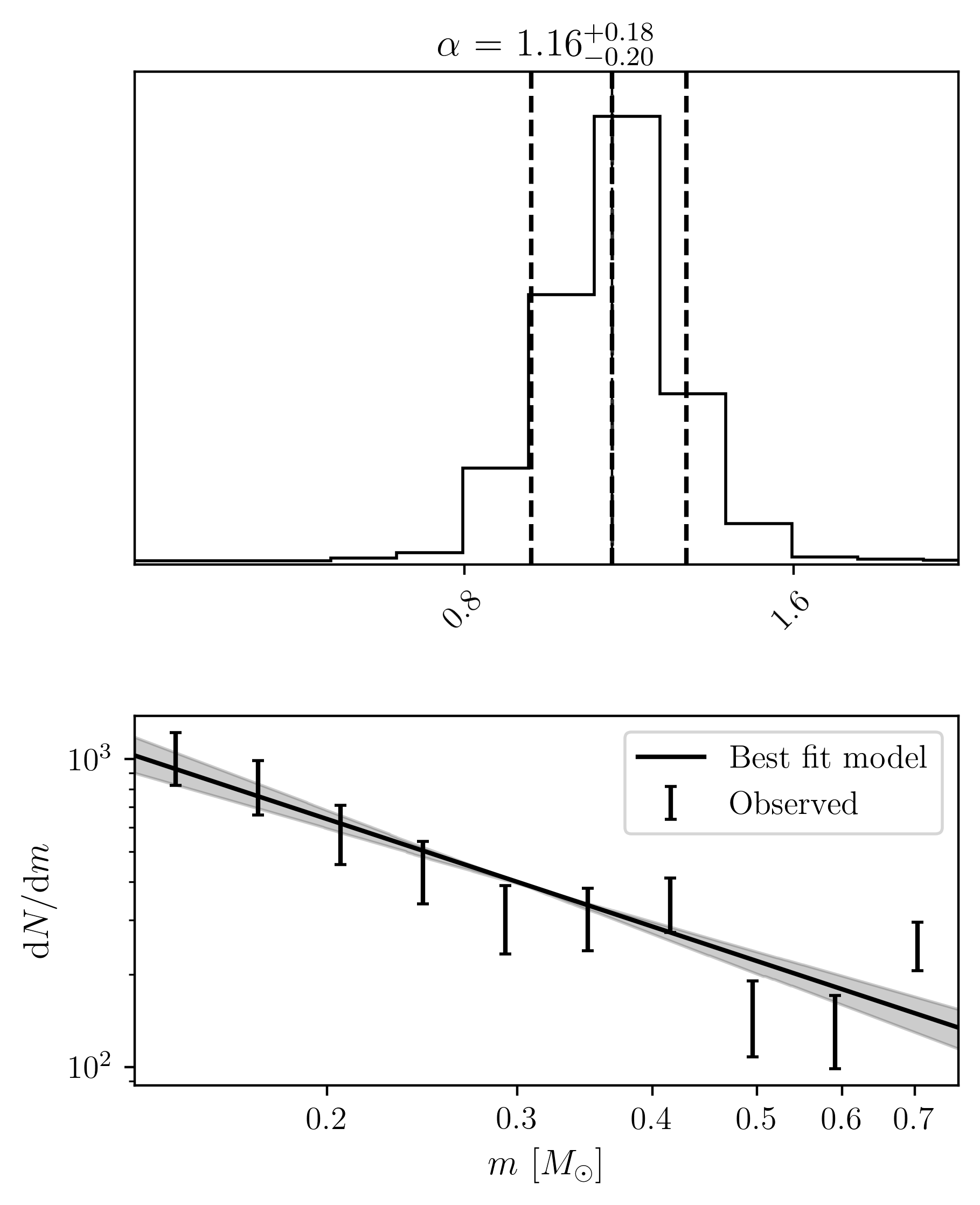}
    \caption{Mass function slope obtained from the HST catalogue. The \textit{top panel} shows the posterior distribution for the single power law parameter $\alpha$ obtained using \texttt{Dynesty} sampling on the unbinned data. The dashed lines show the posterior median and confidence interval. Note that the parameter $\alpha$ has not been corrected to account for binary stars yet. In the \textit{lower panel}, we show the single power law fit with its $68\%$ interval (solid line and shaded region). The bins are only defined to visually verify the consistency of the fit with the observations and have been corrected for incompleteness.}
    \label{fig:mass_function_piecewise}
\end{figure}

The Salpeter mass function \citep{Salpeter:1955} was originally used to describe the mass distribution of a stellar population. It was later refined by the Kroupa broken power law \citep{Kroupa:2001} and the Chabrier log-normal distribution \citep{Chabrier:2003}. In this work, we model the stellar distribution with a single power law: \[p(m)  \propto m^{-\alpha}, \quad m > M_{min}, \quad\alpha > 0,\] where $M_{min}$ is the minimum observed mass and $\alpha$ is the slope of the mass function. We define the corresponding normalised likelihood (including completeness corrections) and perform a Bayesian fit using nested sampling \citep[python package \texttt{Dynesty,}][]{Speagle:2020, Koposov:2025} to obtain the posterior distribution of $\alpha$. We then derive the value of $\alpha$ and its uncertainty from the median and the $16$ and $84\%$ credible intervals from the posterior distribution. The top panel of Fig.~\ref{fig:mass_function_piecewise} illustrates the posterior distribution of the slope coefficient of the power law for the HST catalogue, while the bottom panel shows the Salpeter-like single power law fit with its uncertainty. The bins represent the observations corrected for incompleteness; these bins are shown for illustration purposes only, as the fit is performed on the entire dataset independently of binning. 

To account for biases introduced by unresolved binaries, observational uncertainties, and photometric blends, we adopt the following Monte Carlo approach. In a nutshell, we generate $10,000$ synthetic stellar populations with a known mass-function slope $\alpha_{\rm true}$, binary fraction $f_{\rm bin}$, and mass ratios drawn from a flat distribution. Photometry is assigned via isochrone interpolation, accounting for observational uncertainties and blends derived from artificial-star tests. The mass-function slope is then re-measured assuming all stars are single, reproducing the bias present in the observations. We train a Gaussian Process (GP) emulator on these simulations to map the biased slope to the true value as a function of $f_{\rm bin}$ and $N_{\star}$. To propagate the full uncertainty budget, we draw $10,000$ Monte Carlo samples from the observed posteriors of $\alpha$ and $f_{\rm bin}$, evaluate the emulator at each realisation, and sample from the GP predictive distribution, thereby marginalising over both the observational uncertainties and the emulator prediction uncertainty. We also test the sensitivity of the correction to the assumed mass-ratio distribution by repeating the simulations with mass-ratio distribution $p(q)\propto q^{3}$ (favouring equal-mass binaries) and $p(q)\propto q^{-1}$ (favouring low mass ratios). Applying the mass-function slope correction for a flat mass-ratio distribution yields $\alpha_{\mathrm {HST}} = 1.53\pm0.26$.

This method is applied independently to the HST and JWST catalogues (see Appendix~\ref{app:jwst_figs}). In the JWST catalogue, we obtain masses ranging from $0.10$ to $0.77\ M_\odot$, which is very similar to the mass range obtained in the HST catalogue. The corrected mass slope accounting for binaries is $\alpha_{\mathrm{JWST}} = 1.28 \pm 0.21$; it varies between $1.2$ and $1.3$ depending on the mass-ratio distribution profile. The final mass function slopes for the two catalogues are listed in Table~\ref{tab:salpeter}, along with a comparison to the Salpeter slope.

\begin{table}
  \centering
  \caption{Parameters of the single power law model to fit the IMF from HST and JWST data.}
  \begin{tabular}{lccc}
    \hline\hline
    Parameter & Salpeter & HST Fit & JWST Fit \\
    \hline
    $\alpha$ before binary correction & ... & $1.16^{+0.18}_{-0.20}$ & $0.98^{+0.16}_{-0.16}$
    \\[1.2ex]
    $\alpha$ after binary correction & 2.35 & $1.53\pm0.26$ & $1.28 \pm 0.21$ \\
    \hline
  \end{tabular}
  \label{tab:salpeter}
\end{table}

In a second step, we attempt to model the stellar mass distribution with a broken power law \citep{Kroupa:2001}: \[
\left\{
\begin{aligned}
p_{\mathrm{low}}(m)  &\propto m^{-\alpha_{\mathrm{low}}},  && M_{min} \leq m \leq M_{\mathrm{break}}, &&\alpha_{low} > 0 \\
p_{\mathrm{high}}(m) &\propto m^{-\alpha_{\mathrm{high}}}, && m \geq M_{\mathrm{break}}, &&\alpha_{\mathrm{high}} > 0
\end{aligned}
\right.
\] Here $\alpha_{\mathrm{low}}$ and $\alpha_{\mathrm{high}}$ are the slopes of the mass function for low- and high-mass stars, respectively. We infer the parameters $M_{\mathrm{break}}$, $\alpha_{\mathrm{low}}$ and $\alpha_{\mathrm{high}}$ by sampling the posterior distribution with the \texttt{emcee} package \citep{Foreman-Mackey:2013}.We fit this model independently to the HST and JWST catalogues. The resulting parameters are compared to the Kroupa broken power law in Table~\ref{tab:imf_fits}. However, our high-mass sample is limited: we have to constrain the mass break with a strong prior, and the high-mass slope has large uncertainties. Hence, the single power law model is better suited for a system like Dra\,II.

\begin{table}
  \centering
  \caption{Parameters of the broken power law model to fit the IMF from HST and JWST data.}
  \begin{tabular}{lccc}
    \hline\hline
    Parameter & Kroupa & HST Fit & JWST Fit \\
    \hline
    Break mass $M_{\mathrm{break}}$ ($M_\odot$) & 0.50 & $0.51_{-0.05}^{+0.05}$ & $0.52_{-0.04}^{+0.05}$  \\
    [1.2ex]
    Low-mass slope $\alpha_{\mathrm{low}}$ & 1.3 & $1.21_{-0.15}^{+0.15}$ & $0.93_{-0.13}^{+0.13}$  \\
    [1.2ex]
    High-mass slope $\alpha_{\mathrm{high}}$ & 2.3 & $1.35_{-0.42}^{+0.45}$ & $1.91_{-0.44}^{+0.45}$ \\
    \hline
  \end{tabular}
  \label{tab:imf_fits}
\end{table}

\begin{figure}
    \centering
    \includegraphics[width=\columnwidth]{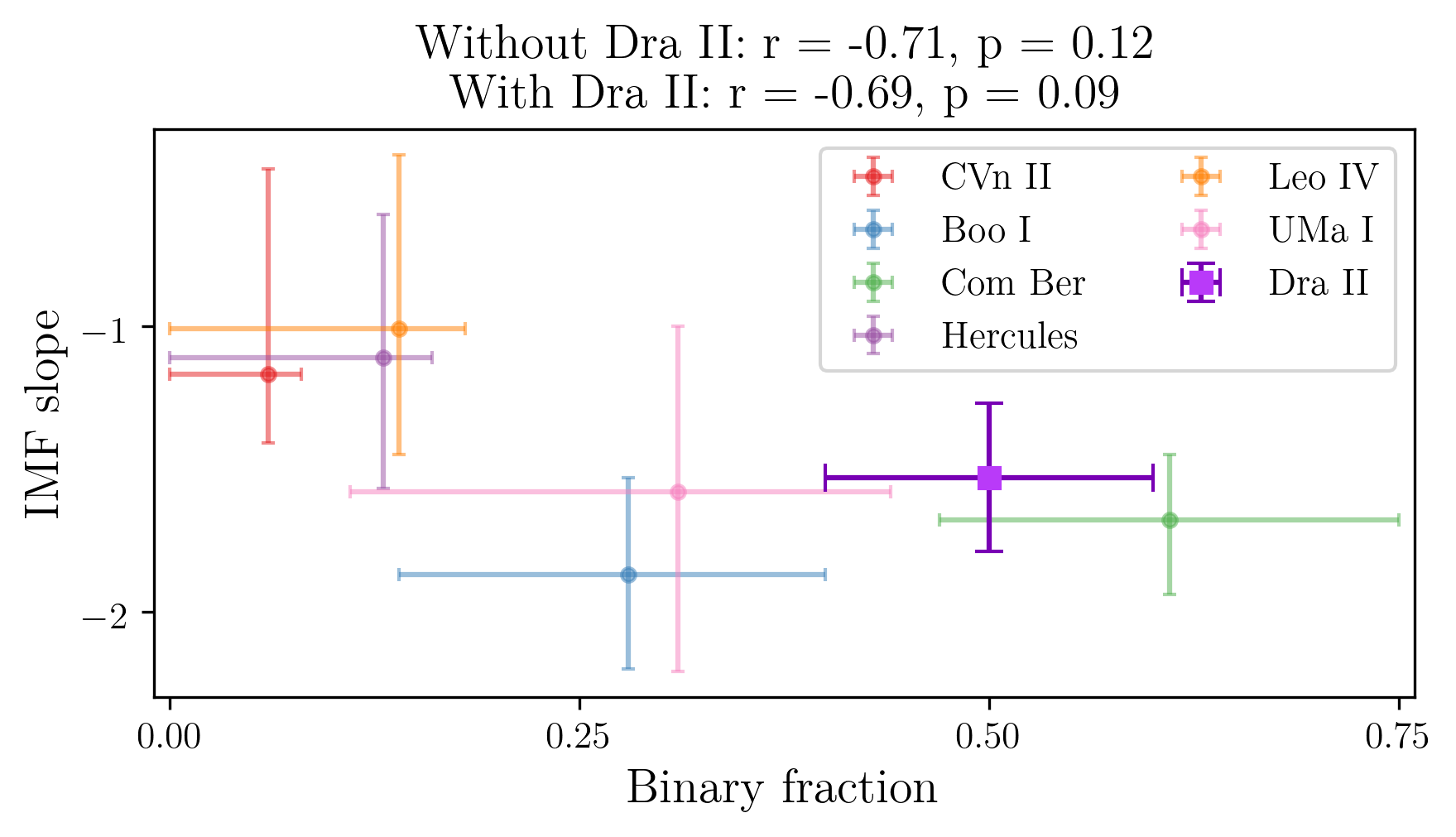}
    \caption{Correlation between IMF slope and binary fraction, reproduced using the values in Table~4 of \citet{Gennaro:2018b}, for six Milky Way UFDs and Dra\,II (added as a purple square). Dra\,II is consistent with the anticorrelation between IMF and binary fraction.}
    \label{fig:imf_slope_binary}
\end{figure}

In addition, in the absence of strong and clearly characterised dynamical evolution, the present-day mass function may approximate the IMF, allowing comparison with other systems. So, using our flat mass-ratio binary fraction ($f_{\rm bin}^{\rm total} = 0.5 \pm 0.1$), Dra\,II is also consistent with the empirical anticorrelation between IMF and binary fraction identified by \citet{Gennaro:2018b} across Milky Way UFDs (Canes Venatici II, Boötes I, Coma Berenices, Hercules, Leo IV, and Ursa Major I; see Fig.~\ref{fig:imf_slope_binary}). We compute the Pearson correlation coefficient $r$, and the corresponding $p$-value, defined as the probability of obtaining a similar or higher value of $|r|$ from a sample with the same number of data but no correlation. We evaluate these quantities both with and without Dra\,II. Including Dra\,II changes $r$ only marginally, from $-0.71$ to $-0.69$, while the $p$-value decreases from $0.12$ to $0.09$. The negative trend is therefore consistent with the inclusion of Dra\,II (shown as the purple square); its addition does not substantially alter the inferred correlation. Dra\,II provides an additional data point in the high binary-fraction regime, reinforcing the trend previously suggested by Coma Berenices. Dra\,II is therefore consistent with the previously reported behaviour, although the physical origin of this relation remains unclear.

\subsection{Mass segregation}

We investigate the presence of mass segregation in Dra\,II, a phenomenon commonly observed in globular clusters but generally not in dwarf galaxies. Whether Dra\,II exhibits mass segregation remains debated \citep{Longeard:2018, Baumgardt:2022}. Our analysis is based on a photometric sample that covers a more central region than the spectroscopic sample of \citet{Longeard:2018}, and that applies an additional proper‑motion selection compared to the photometric study of \citet{Baumgardt:2022} to identify Dra\,II members and reduce contamination from faint foreground and background sources. We include binary stars in this analysis, acknowledging that this may introduce some bias; however, because we identify binaries only among the bright stars and not the faint ones, removing binaries from the bright sample but not the faint sample would introduce an additional bias.

\begin{figure}
    \centering
    \includegraphics[width=1\linewidth]{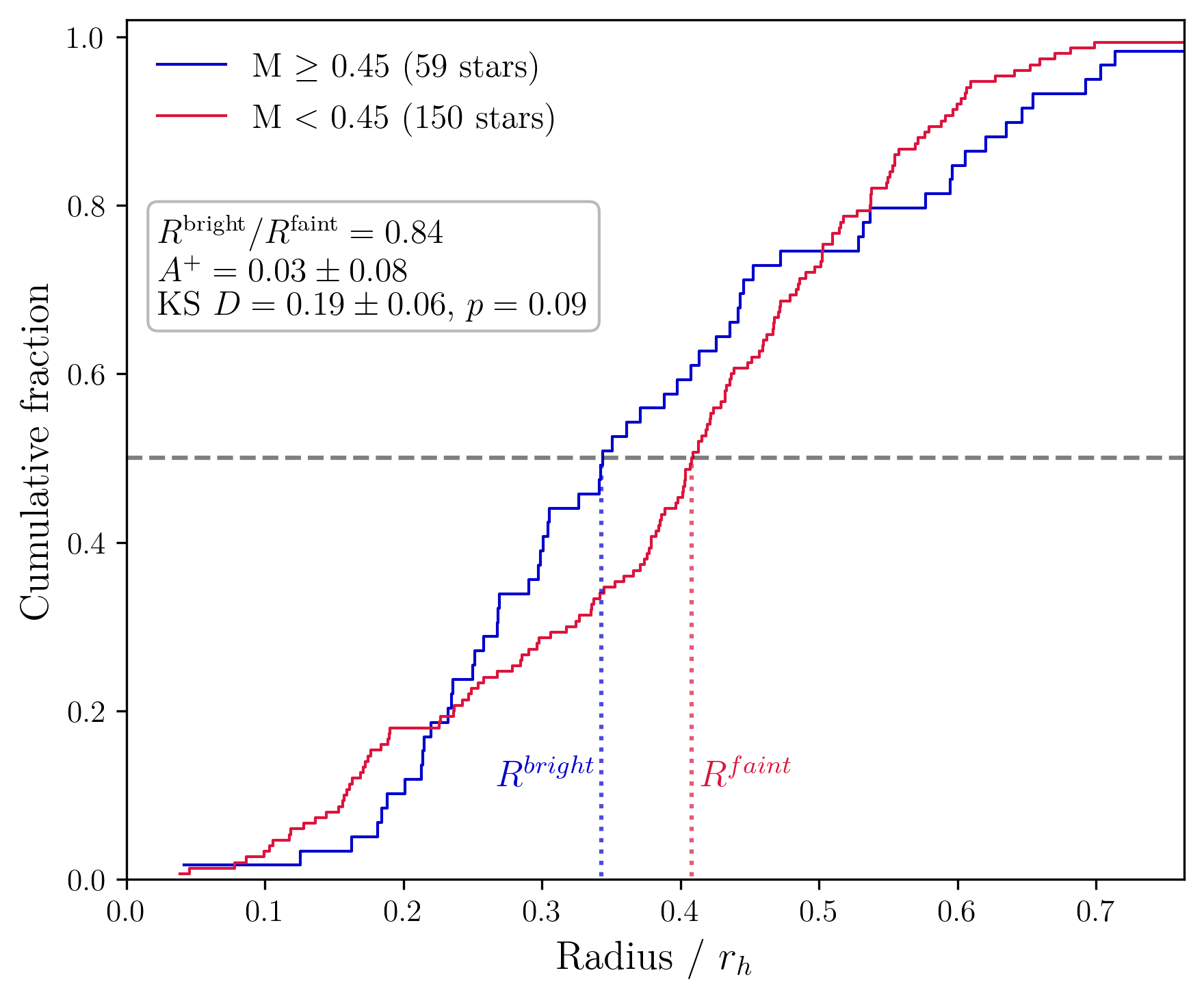}
    \caption{Cumulative radial distribution of Dra\,II stars, split into high-mass (${M \geq 0.45\,M_\odot}$; blue) and low-mass ($M < 0.45\,M_\odot$; red) subsamples. Different quantifiers of the degree of mass segregation are indicated. Mass segregation can be quantified by comparing $R^{\mathrm{bright}}$ and $R^{\mathrm{faint}}$ at a cumulative fraction of 0.5; however, in our case this single measurement does not capture the overall trend.}
    \label{fig:mass_seg_cumul}
\end{figure}

A common approach to quantify mass segregation is to compare the cumulative radial distributions of high- and low-mass stars, as done in both previous studies of Dra\,II. However, the inferred degree of segregation depends on the choice of mass threshold separating the two populations. This is particularly relevant for Dra\,II, where the small number of stars means that the shape of the cumulative distributions, and hence the inferred ratio, depends strongly on the choice of mass bins. Splitting the sample into two equal-sized populations naturally results in a relatively low threshold because of the large number of faint, low-mass stars and the scarcity of bright, massive stars in Dra\,II; indeed, the median mass of our sample is only $0.3\,M_\odot$. We therefore consider several fixed mass thresholds---$0.2$, $0.45$, and $0.7\,M_\odot$---to divide the sample into high- and low-mass populations. We also split the sample in three mass bins to compare the lowest- and highest-mass populations while excluding the intermediate-mass stars, following an approach similar to \citet{Longeard:2018}. The resulting cumulative radial distributions for a mass threshold of $0.45\,M_\odot$ are shown in Fig.~\ref{fig:mass_seg_cumul}.

One way to quantify the difference between the radial distributions, following \citet{Baumgardt:2022}, is to measure the radii $R^{\rm bright}$ (blue dotted line in Fig.~\ref{fig:mass_seg_cumul}) and $R^{\rm faint}$ (red dotted line) at which the cumulative distribution of the bright and faint populations reaches $0.5$ (horizontal dashed line). We then compute the ratio $R^{\rm bright} / R^{\rm faint}$: a ratio below unity indicates that the bright, more massive stars are more centrally concentrated than the faint, lower-mass stars. However, this ratio is based on a single point of the cumulative distribution and therefore does not reflect differences in the shapes of the two profiles, as clearly illustrated in Fig.~\ref{fig:mass_seg_cumul}.

To capture the difference between the two radial distributions over the full radial range, we additionally use the $A^+$ parameter. Initially defined as the area between the cumulative radial distributions of two populations \citep{Alessandrini:2016, Lanzoni:2016}, the $A^+$ parameter was used to quantify the relative spatial concentration of blue straggler stars with respect to a reference population in star clusters. It has subsequently been applied more broadly to compare the spatial distributions of different stellar populations in globular clusters \citep{Dalessandro:2019}; in our case, we define it as:

\[A^{+} = \int_{R_{\min}}^{R_{\max}}
\left[\phi_{\rm high\,M}(r) - \phi_{\rm low\,M}(r)\right]\,{\rm d}r,\]

where $\phi_{\rm high\,M}(r)$ and $\phi_{\rm low\,M}(r)$ are the cumulative radial distributions of the high- and low-mass populations, respectively, normalised to unity at $R_{\max}$. With this definition, positive values indicate that the high-mass population is, on average, more centrally concentrated than the low-mass population. Conversely, negative values indicate inverse mass segregation, with the high-mass stars preferentially located at larger radii. Although $A^+$ accounts for the cumulative difference between the two distributions over the full radial range, it remains a single integrated measure and can therefore obscure local differences: positive and negative differences between the two cumulative distributions at different radii can partially cancel, resulting in an $A^+$ value close to zero despite significant differences between the populations over specific radial ranges.

As a complementary measure, we perform a two-sample Kolmogorov-Smirnov (KS) test to assess whether the radial distributions of the high- and low-mass populations are statistically distinguishable.

For a mass threshold of $0.45M_\odot$ separating the low- and high-mass samples, we find $R^{\rm bright}/R^{\rm faint} = 0.84$, $A^+ = 0.03 \pm 0.08$, and $KS = 0.19 \pm 0.06$ ($p=0.09$). These values vary with the different mass thresholds, illustrating the importance of the choice of mass bins; however, like for the threshold $0.45M_\odot$, none of the indicators provide strong evidence for mass segregation in Dra\,II. Besides, the $A^+$ and $KS$ indicators illustrate that the ratio $R^{\rm bright}/R^{\rm faint}$ does not fully capture the spatial distribution, especially for a system with such a small number of stars, where the cumulative distribution exhibits larger fluctuations. Moreover, we apply our mass-segregation analysis to the HST, JWST, and combined catalogues. The JWST data contain a gap between the two NIRCam detectors resulting in an incomplete coverage of the central region (see Fig.~\ref{fig:HST_JWST_FoVs}). To mitigate this, we considered two approaches: adjusting the stellar coordinates to account for the missing region, and performing the analysis separately within each detector's field of view.

To further assess the possible presence of mass segregation, we apply an independent method based on minimum spanning trees \citep[MSTs;][]{Allison:2009}, implemented using the \texttt{MiSTree} package \citep{Naidoo:2019}. This method has several advantages: it does not require to specify the centre of the stellar system, it does not rely on a predefined mass threshold, and it provides a quantitative measure of the degree of mass segregation. The MST analysis likewise finds no significant evidence for mass segregation. Given the limited sample size, the results are subject to substantial statistical fluctuations; but all approaches and catalogues yield consistent results, indicating that there is no strong evidence of mass segregation in Dra\,II.

\section{Summary and discussion}
\label{sec:discussion}

We have applied state-of-the-art data reduction codes to the multi-band imaging of Draco\,II obtained with HST and JWST. Dra\,II members are selected based on their proper motion, measured over a seven-year baseline, and on photometric quality criteria. The detailed study of the resulting deep photometry reveals key features of Dra\,II. The CMDs show a well-defined MS reaching F814W $ = 27$\,mag, a turn-off point at F814W $ \approx 20$\,mag, and a few brighter stars. Isochrone fitting of the MS of Dra\,II indicates that Dra\,II is a $13.5_{-1.5}^{+1.7}$\,Gyr old, $\alpha$-enhanced ($\afe$ = +0.4\,dex) system, located at a heliocentric distance of $19.2\pm0.9$\,kpc. The absence of evidence for a significant age range is consistent with a rapid star formation episode, typical of UFDs \citep{Brown:2014, Simon:2019}. This bursty star formation would also explain the $\alpha$-enhanced stellar population \citep{Tinsley:1979}. The average metallicity inferred from the isochrone is $\feh_{CMD} = -2.5 \pm0.2$\,dex. The metallicity and $\alpha$-enhancement of Dra\,II are consistent with the $\afe$--$\feh$ correlation found in more luminous dwarf galaxies \citep{Vargas:2013}. However, due to the sparsely populated turn-off region, the parameters derived from isochrone fitting remain partially degenerate.

In the F814W vs F606W$-$F814W and F150W vs F090W$-$F150W CMDs, we identify for the first time several unresolved binary stars in Dra\,II, for different minimum mass ratios $q_{\rm lim}$. The identification of unresolved binary stars is critical because they affect the population characteristics and structural properties inferred from the photometric catalogues. From the HST catalogue, we measure $f_{bin}^{q>0.5} = 0.21 \pm 0.07$. Binary fractions derived for the JWST catalogue are slightly higher but remain within $1\sigma$ ($1.5\sigma$ for $q_{\rm lim} = 0.6$) of the HST results. With Dra\,II's total absolute magnitude of $M_V \approx -2.9$, a purely baryonic star cluster would be expected to have $f_{bin}^{q > 0.5} \approx 0.15$ \citep{Milone:2012}. On the other hand, environments with a lower density of stars, such as open clusters or dwarf galaxies, retain higher binary fractions because their binary systems are less likely to be disrupted. Our measured value is significantly higher than the typical GC expectations, and near the upper range of values observed in open clusters \citep{Cordoni:2023}, supporting the classification of Dra\,II as a low stellar density environment. If Dra\,II is a dwarf galaxy, the binary fraction obtained in this work would support that dwarf galaxies host higher binary fractions than star clusters. Moreover, the high binary fraction is also consistent with empirical trends showing that more metal-poor systems tend to host larger binary populations \citep{Moe:2019, Sana:2025}. Finally, we identify a potential blue straggler, which could be the product of the merger of a binary system.

Previous works applied different methods to investigate the metallicity of Dra\,II. Isochrone fitting provides an initial estimate of the system's mean metallicity, but does not constrain any intrinsic spread. \citet{Martin:2016} first suggested a possible metallicity spread in Dra\,II. \citet{Longeard:2018} refined the measurement of the average metallicity with spectroscopy and set an upper bound on the metallicity spread based on the Pristine survey CaHK photometry, which is sensitive to metallicity. \citet{Fu:2023} quantified this spread using the HST filters F814W, F475W and F395N in a colour–colour diagram analogous to the CaHK diagram. In this work, we reproduce the same colour–colour diagram and infer a median metallicity of $\feh_{\rm CaHK} = -2.38 \pm 0.17$ with a significant metallicity spread of $\sigma_{\feh} = 0.85 \pm 0.13$\,dex, confirming the presence of the metallicity spread identified by \citet{Fu:2023}. Compared to \citet{Fu:2023}, we infer a higher mean metallicity and a larger dispersion. Several factors contribute to this difference. First, although BaSTI and MIST isochrones produce similar tracks in standard CMDs, they differ in the CaHK diagram for $F475W - F814W \gtrsim 1.4$. In this regime, BaSTI isochrones systematically assign higher metallicities than MIST at a given colour and magnitude. To assess the impact of this difference, we split our sample into bright and faint stars based on their $F475W-F814W$ colour. The bright subsample shows a tighter, more metal-poor population, with ${\feh_{\rm CaHK,\ bright} = -2.61 \pm 0.15\,\mathrm{dex}}$ and ${\sigma_{\feh,\rm\ bright} = 0.68 \pm 0.11\,\mathrm{dex}}$, in better agreement with \citet{Fu:2023}. By contrast, the faint stars, which lie in the colour range where BaSTI and MIST differ most, exhibit a broader and more metal-rich distribution, with ${\feh_{\rm CaHK,\ faint} = -1.61 \pm 0.37\,\mathrm{dex}}$ and ${\sigma_{\feh,\rm\ faint} = 0.91 \pm 0.26\,\mathrm{dex}}$. This faint population drives the larger overall spread we measure. In addition, we exclude binary stars from the fit of the metallicity distribution. Including binaries reduces the overall inferred spread from ${\sigma_{\feh} = 0.86}$ to ${\sigma_{\feh} = 0.80\,\mathrm{dex}}$, indicating that our treatment of binaries also contributes to the larger dispersion we report.

One main limitation of the CaHK method is that it can be applied only to the brightest stars, a sample whose size is limited by the faintness of Dra\,II. Hence, we also investigate the metallicity distribution of Dra\,II's fainter stars by verticalizing the CMD and fitting its cumulative distribution function using the Python package \texttt{Dynesty} \citep{Speagle:2020, Koposov:2025}. We confirm its low metallicity ($\feh_{\rm CDF} = -2.34 \pm 0.15$) as well as the presence of a metallicity spread ($\sigma_{\feh} = 0.63 \pm 0.15$). Both bright (CaHK) and faint (CDF) stars exhibit a very low mean metallicity and a substantial intrinsic spread, confirming the presence of a genuine metallicity spread in Dra\,II. The analysis of the faint stars extends this result to a much larger sample and to lower luminosities, where the CaHK method becomes inapplicable. However, the most metal-rich stars identified in the faint CaHK subsample should be interpreted with caution. Their inferred metallicities are particularly sensitive to the choice of isochrones, as BaSTI systematically assigns more metal-rich values than MIST in this regime. While this does not erase the overall spread, it implies that the detailed shape of the metal-rich tail remains uncertain. Taken together, these results confirm that Dra\,II possesses a significant metallicity spread, detected by different methods reaching faint stars, and that this spread is larger than previously reported once fainter stars are included and binaries are excluded. Both the very low metallicity and the substantial metallicity spread of its stellar population support the classification of Dra\,II as a dwarf galaxy.

We use BaSTI isochrones to interpolate stellar masses for Dra\,II member stars and construct the corresponding present-day mass functions. A single power-law fit yields slopes of $\alpha = 1.53 \pm 0.26$ for the HST catalogue and $1.28 \pm 0.21$ for the JWST catalogue, both significantly flatter than a Salpeter IMF. Dra\,II contains only a small stellar population, so statistical fluctuations may partially account for any deviation from the Salpeter value. To explore possible deviations from a single-slope description, we also fit a broken power-law model. For the HST catalogue, we obtain a break mass of $M_{\rm break} = 0.51 \pm 0.05\,M_\odot$, with a low-mass slope of $\alpha_{\rm low} = 1.21 \pm 0.15$ and a high-mass slope of $\alpha_{\rm high} = 1.35 \pm 0.45$ (see Table~\ref{tab:imf_fits}). The single-slope fit yields a slope shallower than the Salpeter value. While the low-mass slope of the broken power-law fit is broadly consistent with the commonly accepted Kroupa values, the high-mass slope remains poorly constrained because only a small number of stars lie above the break mass. For this reason, a single power-law model provides a more robust description of the stellar mass distribution. Moreover, we compare the spatial distribution of bright and faint Dra\,II stars but find no strong evidence supporting that this system is mass segregated.

In the absence of strong and clearly characterised dynamical evolution, the present-day mass function may approximate the IMF, allowing comparison with other systems. Our results are consistent with previous studies of ultra-faint systems. IMF slopes of $0.8-1.3$ have been reported for GCs and DGs \citep{Geha:2013, Gennaro:2018b, Baumgardt:2022}; while \citet{Gennaro:2018} measure $\alpha_{low} = 1.18_{-0.33}^{+0.49}$, $\alpha_{high} = 1.88_{-0.49}^{+0.43}$ and $M_{break} = 0.57_{-0.08}^{+0.12}$ for the dwarf galaxy Coma Berenices. Dra\,II is also consistent with the empirical anticorrelation between IMF and binary fraction identified by \citet{Gennaro:2018b} across Milky Way UFDs, as illustrated in Fig.~\ref{fig:imf_slope_binary}.

In summary, the deep HST and JWST photometry of Draco\,II reveals a $13.5$\,Gyr old, $\alpha$-enhanced ($\afe = +0.4$\,dex), and extremely metal-poor ($\feh_{CaHK} = -2.38 \pm 0.17$\,dex) stellar system with a substantial intrinsic metallicity dispersion ($\sigma_{\feh} = 0.85 \pm 0.13$\,dex), a high unresolved binary fraction ($f_{bin}^{q>0.5} = 0.21 \pm 0.07$), and a stellar mass function consistent with those observed in other ultra-faint dwarf galaxies. We find no evidence supporting the presence of mass segregation. The combination of its chemical properties, possible metallicity spread, dynamical mass estimates, and low present-day luminosity suggests that Dra\,II may originally have been significantly more massive, although the evolutionary processes responsible for its present state remain uncertain. Altogether, these results favour a classification of Dra\,II as an ultra-faint dwarf galaxy rather than a star cluster. Draco II can therefore serve as a key benchmark for both Galactic Archaeology and $\Lambda$CDM cosmology. As an ultra-faint dwarf galaxy, it can be used to probe galaxy formation at the faintest luminosities and the properties of dark matter in the smallest bound halos.

\section*{Acknowledgements}
We thank the anonymous referee for their constructive input. We thank S. Cassisi for making metal-poor $\alpha$-enhanced BaSTI isochrones available for our analysis. All of the data presented in this paper were obtained from the Mikulski Archive for Space Telescopes (MAST) at the Space Telescope Science Institute. STScI is operated by the Association of Universities for Research in Astronomy, Inc., under NASA contract NAS5-26555 for HST and NASA contract NAS5-03127 for JWST. Support to MAST for these data is provided by the NASA Office of Space Science via grand NAG57584 and by other grants and contracts.

\textit{Facilities:} HST (ACS), HST (WFC3), JWST (NIRCAM)

\textit{Software:} \texttt{KS2} \citep{Anderson:2008}, \texttt{astropy} \citep{AstropyCollaboration:2018}, \texttt{NumPy} \citep{Oliphant:2006}, \texttt{matplotlib} \citep{Hunter:2007}, \texttt{emcee} \citep{Foreman-Mackey:2013}, \texttt{Dynesty} \citep{Speagle:2020, Koposov:2025}, \texttt{MiSTree} \citep{Naidoo:2019}.

\section*{Data Availability}
The data underlying this article will be shared on reasonable request to the corresponding author.






\bibliographystyle{mnras}
\bibliography{biblio} 



\appendix

\section{Testing different values of \texorpdfstring{$\afe$}{[alpha/Fe]}}
\label{app:alpha_tests}

We chose in this paper to always use $\alpha$-enhanced isochrones ($\afe = +0.4$\ dex), as the $\alpha$-enhancement distribution of Dra\,II stars is unknown. However this assumption is questionable, hence we reproduce some analyses here using isochrones having ${\afe = 0.0\,\mathrm{dex}}$ to verify whether our assumption significantly impacts our results. 

\begin{figure}
    \centering
    \includegraphics[width=1\linewidth]{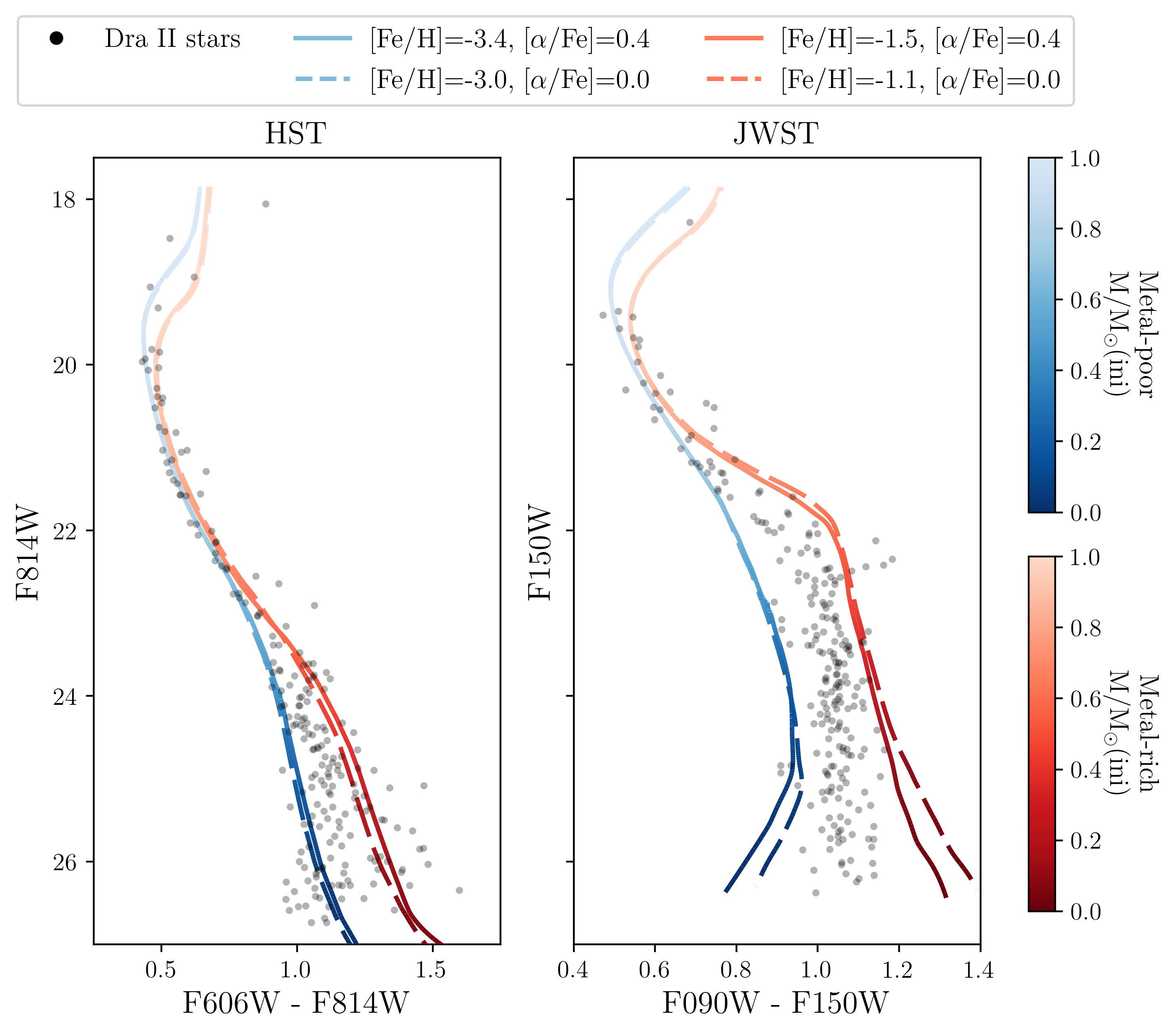}
    \caption{HST and JWST CMDs of Dra\,II, with four isochrones: two $\alpha$-enhanced, one metal-poor and one metal-rich, and two with solar $\alpha$, one metal-poor and one metal-rich.}
    \label{fig:app_alpha_tests_cmd}
\end{figure}

First, we compare isochrone tracks in the CMDs of Dra\,II. Fig.~\ref{fig:app_alpha_tests_cmd} illustrates both HST (left panel) and JWST (right) CMDs. We show four isochrones in these CMDs: two metal-rich (red), and two metal-poor (blue). For each metallicity, one has $\afe = 0.0$ (dashed) and the other is $\alpha$-enhanced (solid line), with a lower metallicity to match best the solar-alpha isochrone in the CMD. The four isochrones are colour-coded by the initial mass of stars on the MS. This figure illustrates two elements: (i) for both metal-rich and metal-poor isochrones, there is an $\alpha$-enhanced isochrone that closely matches the solar-alpha isochrone, with only minor differences at the faint end of the CMD; and (ii) the masses of Dra\,II stars, which are extracted from the position of stars in the CMD compared to a grid of isochrones, do not appear to depend strongly on the $\alpha$-enhancement of the isochrones chosen, i.e. the colour-coding of the $\alpha$-enhanced and the solar-$\alpha$ isochrones is similar for a given F814W or F150W magnitude. Finally, although we have access to a fine grid in $\feh$ (in steps of $0.1$\,dex), we can only obtain isochrones with $\afe = 0.0$ or $\afe = 0.4$\,dex, which limits our ability to constrain a distribution of $\alpha$-enhancement values in the stellar population of Dra\,II.

\begin{figure}
    \centering
    \includegraphics[width=1\linewidth]{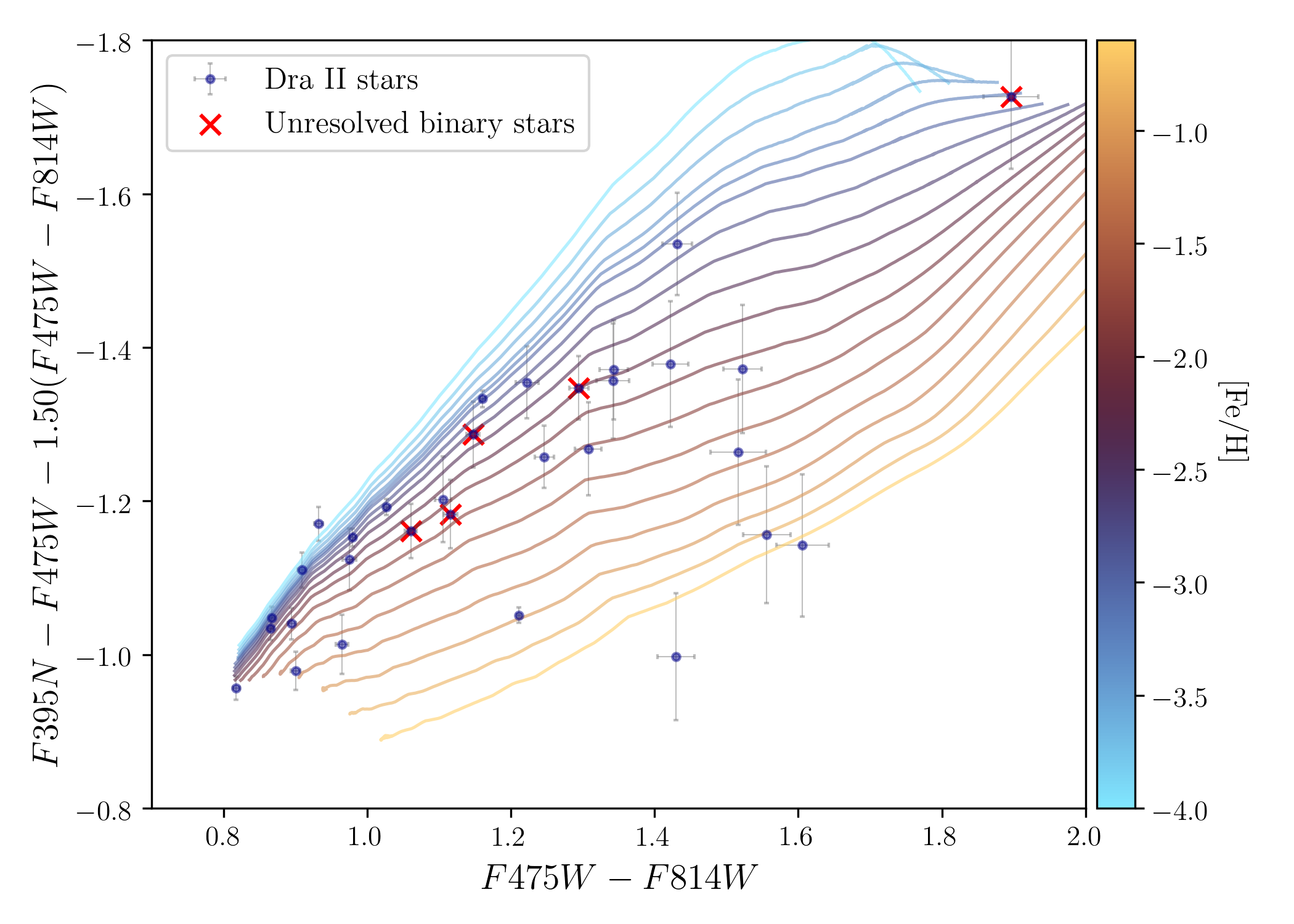}
    \caption{CaHK colour-colour diagram inspired by the Pristine survey \citep{Starkenburg:2017}. Dra\,II stars are the blue points, the identified unresolved binaries are the red crosses. The faintest star has $\mathrm{F814W} = 23$\ mag. A grid of BaSTI isochrones is overplotted, for different metallicities, this time with $\afe=0.0$\ dex.}
    \label{fig:app_cahk_alpha}
\end{figure}

\begin{figure}
    \centering
    \includegraphics[width=0.9\linewidth]{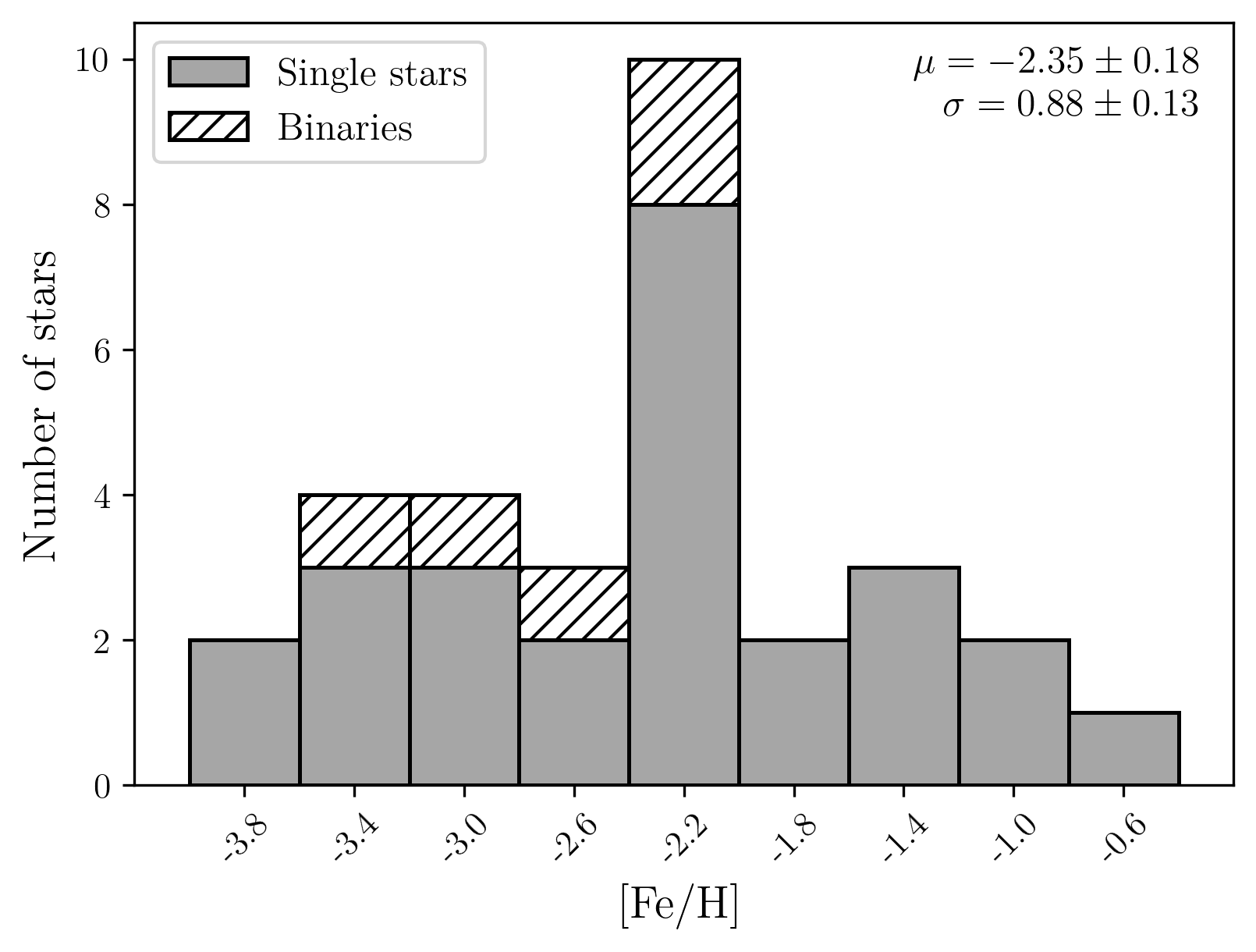}
    \caption{Metallicity distribution from the CaHK analysis using a grid of isochrones with solar $\afe$. The resulting median and standard deviation of the distribution are $\feh = -2.35 \pm 0.18$ and $\sigma_{\feh} = 0.88 \pm 0.13$.}
    \label{fig:app_alpha_result_met}
\end{figure}

We also reproduce our CaHK analysis to obtain the metallicity distribution of bright stars, as explained in Section~\ref{sec:cahk}. We use a grid of isochrones with the same parameters ($\feh$ varying between $-4.0$ and $-0.5$ in steps of $0.1$\ dex, age of $13.5$\ Gyr) but this time, with $\afe = 0.0$ instead of $0.4$\ dex. We correct the bias as previously. The resulting CaHK diagram is illustrated in Fig.~\ref{fig:app_cahk_alpha}. Applying the same method, we obtain the metallicity distribution in Fig.~\ref{fig:app_alpha_result_met}. Using the same procedure as before, we obtain $\feh_{CaHK,sol} = -2.34 \pm 0.17$ and $\sigma_{\feh} = 0.86 \pm 0.12$. For comparison, when using the grid of $\alpha$-enhanced isochrones with $\afe=0.4$\ dex, we found $\feh_{CaHK} = -2.35 \pm 0.18$\,dex and $\sigma_{\feh} = 0.88 \pm 0.13$\,dex. Choosing $\alpha$-enhanced isochrones therefore leads to a slightly lower mean metallicity and a somewhat narrower metallicity distribution, although the two estimates remain consistent within the uncertainties.

\section{Figures for JWST analysis}
\label{app:jwst_figs}

\begin{figure}
    \centering
    \includegraphics[width=1\linewidth]{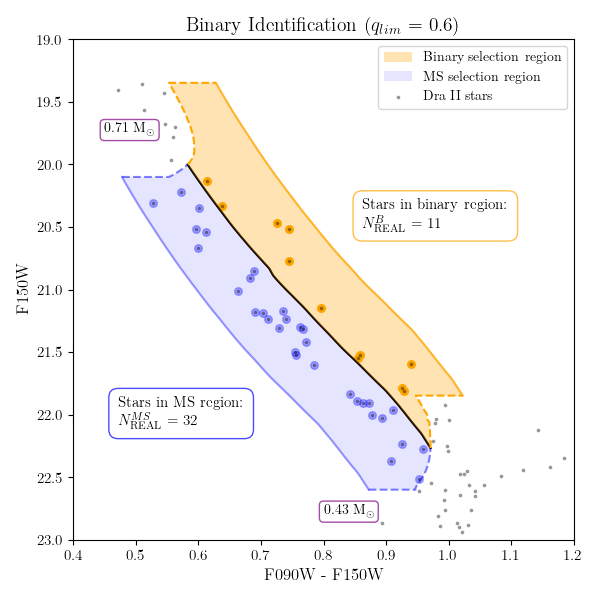}
    \caption{Same as left panel of Fig.~\ref{fig:binaries_all}. Identification of unresolved binary stars for $q_{\mathrm{lim}} = 0.6$ in the JWST catalogue.}
    \label{fig:jwst_bins}
\end{figure}

We present the results obtained from the JWST catalogue. We apply the same method as for the HST data to identify unresolved binary stars, this time using the F150W vs F090W$-$F150W CMD. The magnitude limits are $20.1\leq\mathrm{F150W}\leq 22.6$, corresponding to a stellar mass range of $0.43$ to $0.71\ M_\odot$. The result for $q_{\lim} = 0.6$ is illustrated in Fig.~\ref{fig:jwst_bins}.

\begin{figure}
    \centering
    \includegraphics[width=0.9\columnwidth]{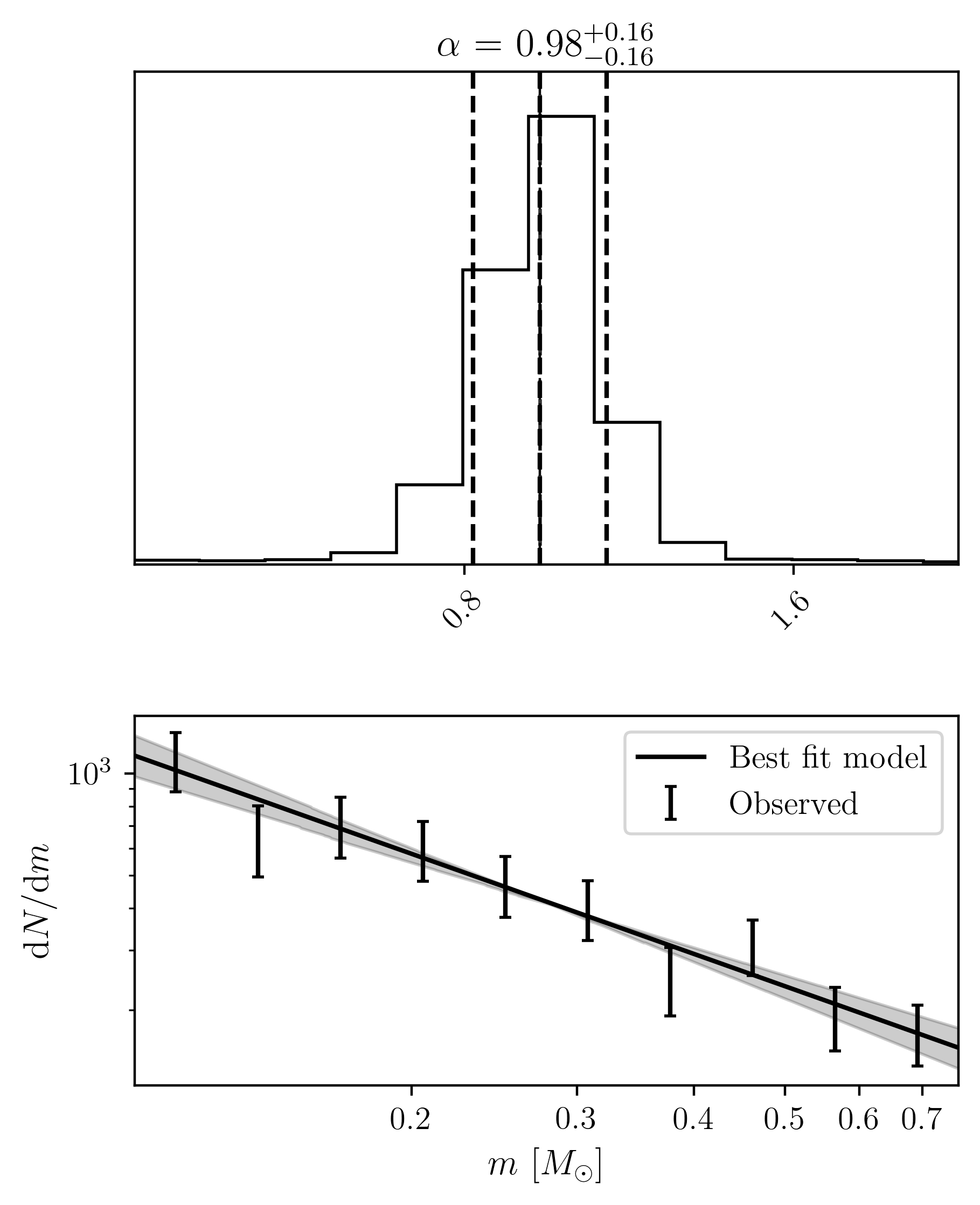}
    \caption{Same as Fig.~\ref{fig:mass_function_piecewise}. Mass function slope obtained from the JWST catalogue, before the binary correction.}
    \label{fig:jwst_mass_function_piecewise}
\end{figure}

Similarly, we obtain the JWST mass slope using the same method as for HST, i.e. estimating per-star masses from a grid of isochrones and fitting a single-slope model to the data. Fig.~\ref{fig:jwst_mass_function_piecewise} shows the slope obtained before correcting it to account for the effect of binary stars; after correction, we obtain $1.28 \pm 0.21$.

\section{Simulated CMDs with different metallicities}
\label{app:simulated_cmds}

\begin{figure}
    \centering
    \includegraphics[width=1\linewidth]{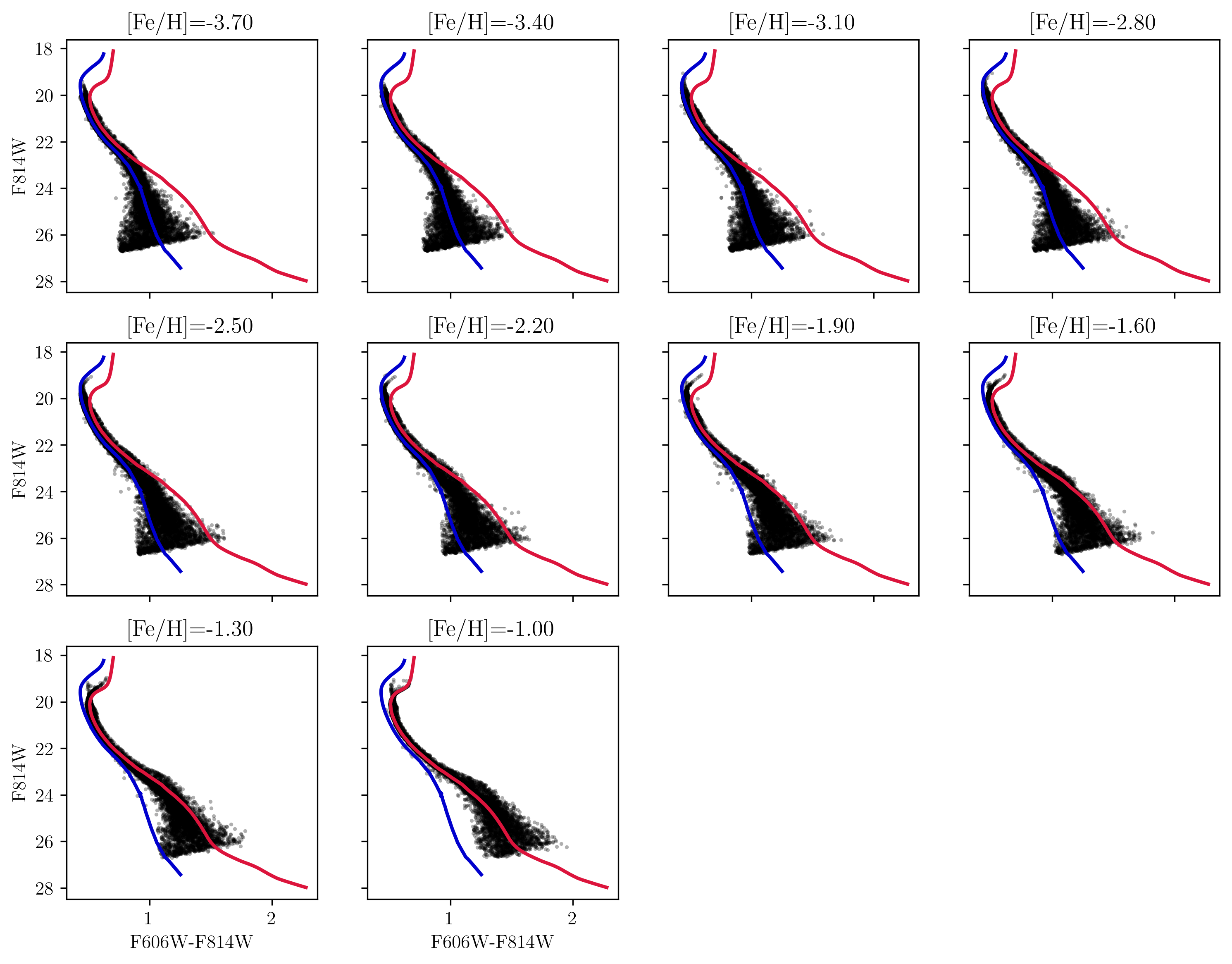}
    \caption{Simulated CMDs of the simple stellar populations with metallicities ranging from $\feh=-3.7$ to $\feh=-1.0$, generated using artificial stars and BaSTI isochrones. They include $30\%$ of binary stars, with a flat mass ratio distribution.}
    \label{fig:app_sim_cmds}
\end{figure}

As described in Section~\ref{sec:dynesty}, we generate a grid of simple stellar populations to be used in our forward modelling framework. These synthetic populations are required to account for photometric uncertainties and blending effects, which are incorporated using the artificial star tests, as well as for the presence of unresolved binary systems. We construct models spanning a metallicity range from $\feh = -3.7$ (blue isochrone in Fig.~\ref{fig:app_sim_cmds}) to $-1.0$ (red isochrone) using BaSTI isochrones. All isochrones assume a fixed age of $13.5$ Gyr, a distance of $19.2$ kpc, and an $\alpha$-enhanced chemical composition of $\afe = +0.4$ dex.

For each isochrone, we include a binary population with a binary fraction of $0.30$ and assume a flat mass-ratio distribution. Photometric scatter and blending effects are applied by sampling the artificial star tests in order to reproduce the observational completeness and measurement uncertainties. The resulting simulated colour–magnitude diagrams are shown in Fig.~\ref{fig:app_sim_cmds}. These illustrate the progressive broadening of the stellar sequences due to photometric uncertainties, as well as the presence of unresolved binaries and a small number of blended sources at all metallicities.

\bsp	
\label{lastpage}
\end{document}